\documentclass{sig-alternate}

\usepackage{amsmath}
\usepackage{paralist}
\usepackage{url}
\usepackage{graphicx}
\usepackage{mathabx}
\usepackage{array}
\usepackage{cite}
\usepackage{pifont}
\usepackage{multicol}
\usepackage{multirow}
\usepackage{makecell}
\usepackage[font=small,skip=4pt]{caption}
\usepackage{xcolor}

\newcounter{SummaryCounter}

\definecolor{gray}{rgb}{0.95,.95,.95}
\definecolor{lightgray}{rgb}{0.5,.5,.5}

\newcommand{\summary}[2]{
\begin{center}
\stepcounter{SummaryCounter}
\noindent\fcolorbox{lightgray}{gray}{
\begin{minipage}{.44\textwidth}
\begin{flushright}\textsc{Summary}\vspace{-0.3cm}\end{flushright}
\emph{#1}\\%[1ex]
\vspace{-0.3cm}
\end{minipage}
}
\end{center}
}

\begin{document}

%\CopyrightYear{2016} 
%\setcopyright{acmcopyright}
%\conferenceinfo{ICSE '16,}{May 14-22, 2016, Austin, TX, USA}
%\isbn{978-1-4503-3900-1/16/05}\acmPrice{\$15.00}
%\doi{http://dx.doi.org/10.1145/2884781.2884793}

\def\BUGSONE{135}
\def\BUGSTWO{323}
\def\PROJECTS{24}
\def\ALGORITHMS{10}

%
% --- Author Metadata here ---
\conferenceinfo{ICSE}{'16 Austin, Texas, USA}
%\CopyrightYear{2007} % Allows default copyright year (20XX) to be over-ridden - IF NEED BE.
%\crdata{0-12345-67-8/90/01}  % Allows default copyright data (0-89791-88-6/97/05) to be over-ridden - IF NEED BE.
% --- End of Author Metadata ---

\title{A Comparison of 10 Sampling Algorithms for\\Configurable Systems}


\numberofauthors{5} %  in this sample file, there are a *total*
\author{
% 1st. author
\alignauthor
Fl\'{a}vio Medeiros\\
       \affaddr{Fed. Univ. of Campina Grande}\\
       \affaddr{Para\'{i}ba, Brazil}
% 2nd. author
\alignauthor
Christian K\"{a}stner\\
       \affaddr{Carnegie Mellon University}\\
       \affaddr{Pittsburgh, Pennsylvania, USA}
% 3rd. author
\alignauthor M\'{a}rcio Ribeiro \\
       \affaddr{Federal University of Alagoas}\\
       \affaddr{Macei\'{o}, Alagoas, Brazil}
\and  % use '\and' if you need 'another row' of author names
% 4th. author
\alignauthor Rohit Gheyi \\
       \affaddr{Fed. Univ. of Campina Grande}\\
       \affaddr{Para\'{i}ba, Brazil}
% 5th. author
\alignauthor Sven Apel\\
       \affaddr{Universit\"{a}t Passau}\\
       \affaddr{Passau, Germany}
}
% There's nothing stopping you putting the seventh, eighth, etc.
% author on the opening page (as the 'third row') but we ask,
% for aesthetic reasons that you place these 'additional authors'
% in the \additional authors block, viz.
%\additionalauthors{Additional authors: John Smith (The Th{\o}rv{\"a}ld Group,
%email: {\texttt{jsmith@affiliation.org}}) and Julius P.~Kumquat
%(The Kumquat Consortium, email: {\texttt{jpkumquat@consortium.net}}).}
% Just remember to make sure that the TOTAL number of authors
% is the number that will appear on the first page PLUS the
% number that will appear in the \additionalauthors section.

\maketitle
\begin{abstract}
%\vspace{-0.15cm}
Almost every software system provides configuration options to tailor the system
to the target platform and application scenario. 
Often, this configurability renders the analysis of every individual system configuration infeasible.
To address this problem, researchers have proposed a diverse set of sampling algorithms.
%, but make certain limiting assumptions: they perform per-file analysis and ignore constraints among configuration options, header files, and build-system information.
We present a comparative study of~\ALGORITHMS~state-of-the-art sampling algorithms regarding their fault-detection capability and size of sample sets. 
The former is important to improve software quality and the latter to reduce the time of analysis.
%Initially, we compared the algorithms accepting some limiting assumptions, but we explicitly evaluate these assumptions in a subsequent study.
In a nutshell, we found that sampling algorithms with larger sample sets are able to detect higher numbers of faults, but simple algorithms with small sample sets, such as \textit{most-enabled-disabled}, are the most efficient in most contexts.
Furthermore, we observed that the limiting assumptions made in previous work influence the number of detected faults, the size of sample sets, and the ranking of algorithms. Finally, we have identified a number of technical challenges when trying to avoid the limiting assumptions, which questions the practicality of certain sampling algorithms.
\end{abstract}

%A category including the fourth, optional field follows...
%\category{D.2.8}{Software Engineering}{Metrics}[complexity measures, performance measures]


%\keywords{Sampling Algorithms, Configurable Systems}

%\vspace{-0.25cm}
%!TEX root = ../main.tex

\section{Introduction}
\label{s:introduction}
\vspace{-0.1cm}

Many software systems can be configured to different hardware platforms, operating systems, and requirements~\cite{Spencer92}.
However, the variability that is inherent to configurable systems challenges quality assurance~\cite{Bosch03,Apel13,Medeiros15,Kuhn04}.
Developers need to consider multiple configurations when they execute tests or perform static analyses to find faults and vulnerabilities.
As the configuration space often explodes exponentially with the number of configuration options, analyzing every individual system configuration becomes infeasible in real-world projects; for example, the \textit{Linux Kernel} has more than 12 thousand compile-time configuration options.
Configuration-related faults that occur only in a subset of all configurations are especially tricky to find~\cite{Medeiros15}.
As such, it is not surprising that many configuration-related faults have been found in highly-configurable systems, such as the \textit{Linux Kernel}, \textit{Gcc}, \textit{BusyBox}, and \textit{Apache}~\cite{Kastner11,Garvin11-2,Tartler12,Medeiros13,Iago14,Medeiros15-3}.

Although researchers have proposed approaches to analyze complete configuration spaces in a sound fashion for some classes of defects~\cite{Thum14,Kastner11,Kastner12-2,Gazzillo12,Tartler12}, the vast majority of mature quality-assurance techniques consider only a single configuration at a time. For example, static-analysis tools operate typically on C code after the C preprocessor has resolved configuration options implemented through conditional compilation (e.g., using \texttt{\#ifdef} directives).
To reuse state-of-the-art tools, such as \emph{gcc}, for detecting configuration-related faults, \textit{sampling} is a viable alternative~\cite{Oster10,Perrouin10,Johansen12,Marijan13,Tartler14}.
That is, instead of analyzing all configurations, one selects a \textit{subset} of configurations to analyze individually. The effectiveness of sampling for detecting configuration-related faults depends significantly on how samples are selected, though.

Several sampling algorithms have been proposed in the literature, such as \textit{t-wise}~\cite{Lei08,Perrouin10,Johansen12,Marijan13}, \textit{statement-coverage}~\cite{Tartler11-2}, and \textit{one-disabled}~\cite{Iago14}.
To select a suitable sampling algorithm, one needs to understand the tradeoffs, especially with regard to effort (i.e., how large are the sample sets) and fault-detection capabilities (i.e., how many faults can be found in the sampled configurations). Unfortunately, a comparison of sampling algorithms for finding configuration-related faults is not available.
More importantly, many proposed sampling algorithms make severe assumptions that may not be realistic for practical applications and that are not always clearly communicated.
For instance, they perform analyses per file instead of globally, and they ignore constraints among configuration options, header files, and build-system information~\cite{Shi05,Kuhn04,Lei08,Nie11}.
Applying sampling algorithms under different assumptions may introduce significant additional effort or reduce coverage, as we will discuss.
A lack of understanding of the tradeoffs and assumptions of sampling algorithms can lead to both undetected faults, which decrease software quality, and time-consuming code analysis, which increases costs.

We conducted a comparative study to analyze~\ALGORITHMS~sampling algorithms in detail to fill that gap. We compared the selected sample sizes and the fault-detection capabilities of the sampling algorithms in a study of~\BUGSONE~known configuration-related faults in 24 open-source C systems, each configurable with conditional compilation. 
Specifically, we analyzed a set of sampling algorithms proposed in the research literature: 5 variations of \textit{t-wise}~\cite{Oster10,Perrouin10,Johansen12,Marijan13}; \textit{statement-coverage}~\cite{Tartler11-2}; \textit{random}; \textit{one-disabled}~\cite{Iago14}; \textit{one-enabled}; and \textit{most-enabled-disabled}.
In summary, we analyzed~\ALGORITHMS~sampling algorithms and 35 combinations of algorithms in two studies.
In the first study, we compared sample sizes and fault-detection capabilities of the different sampling algorithms and their combinations on a large set of open-source systems under favorable assumptions (e.g., ignoring constraints and header files). In the second study, we explored the influence of considering constraints, header files, build-system information, and global analysis, which are often neglected in the literature and practice~\cite{Shi05,Kuhn04,Lei08,Nie11}.

Our results show that all algorithms select configurations with more than 66\% of the configuration-related faults in our corpus. 
Almost 84\% of all faults are detected by enabling or disabling one or two configuration options, but there are also faults that require developers to enable or disable up to \textit{seven} options.
As expected, we found that the algorithms with the largest sample sizes detected the most faults. 
However, simple algorithms with small sample sets, such as \textit{most-enabled-disabled}, are the most efficient in many scenarios.
More interestingly, we identified several novel combinations of algorithms that provide a useful balance between sample size and fault-detection capabilities.

As a further result, we found that considering constraints among configuration options, global analysis, header files, and build-system information influence the performance of most sampling algorithms substantially, up to the point that several algorithms are no longer feasible in practice.
Considering constraints increases the time of analysis significantly, which prohibits us to use some algorithms, such as \textit{three-wise} and \textit{four-wise}, at all. 
Including build-system information increases the size of sample sets slightly, whereas global analysis and analyses that include configuration options from header files turn the analysis to be practically infeasible for most algorithms.

In summary, our main contributions are:
\begin{compactitem}
	\item A comparative study of 10 sampling algorithms and 35 combinations of algorithms regarding their fault-detection capability and size of sample sets;
	\item A study on the influence of considering header files, constraints, build-system information, and global analysis on the performance of sampling algorithms;
	\item A discussion of results showing that some sampling algorithms become infeasible under realistic settings, for example, when incorporating header files and applying global analysis;
	\item A report of significant changes of the efficiency ranking of sampling algorithms when considering different pieces of information, such as build-system and constraint information;
	\item Results supporting sampling algorithms with an efficient balance between sample size and fault-detection capabilities under different assumptions, such as the \textit{most-enabled-disabled} algorithm.
\end{compactitem}

\noindent
All data used in this study are available on our \textit{Website}.\footnote{\scriptsize \url{http://www.dsc.ufcg.edu.br/~spg/sampling/}}

\vspace{-0.4cm}

%!TEX root = ../main.tex

\section{Configuration-Related Faults}
\label{s:motivating}

\begin{figure}[ht]
\centering
\vspace{-0.3cm}
\includegraphics[width=60mm]{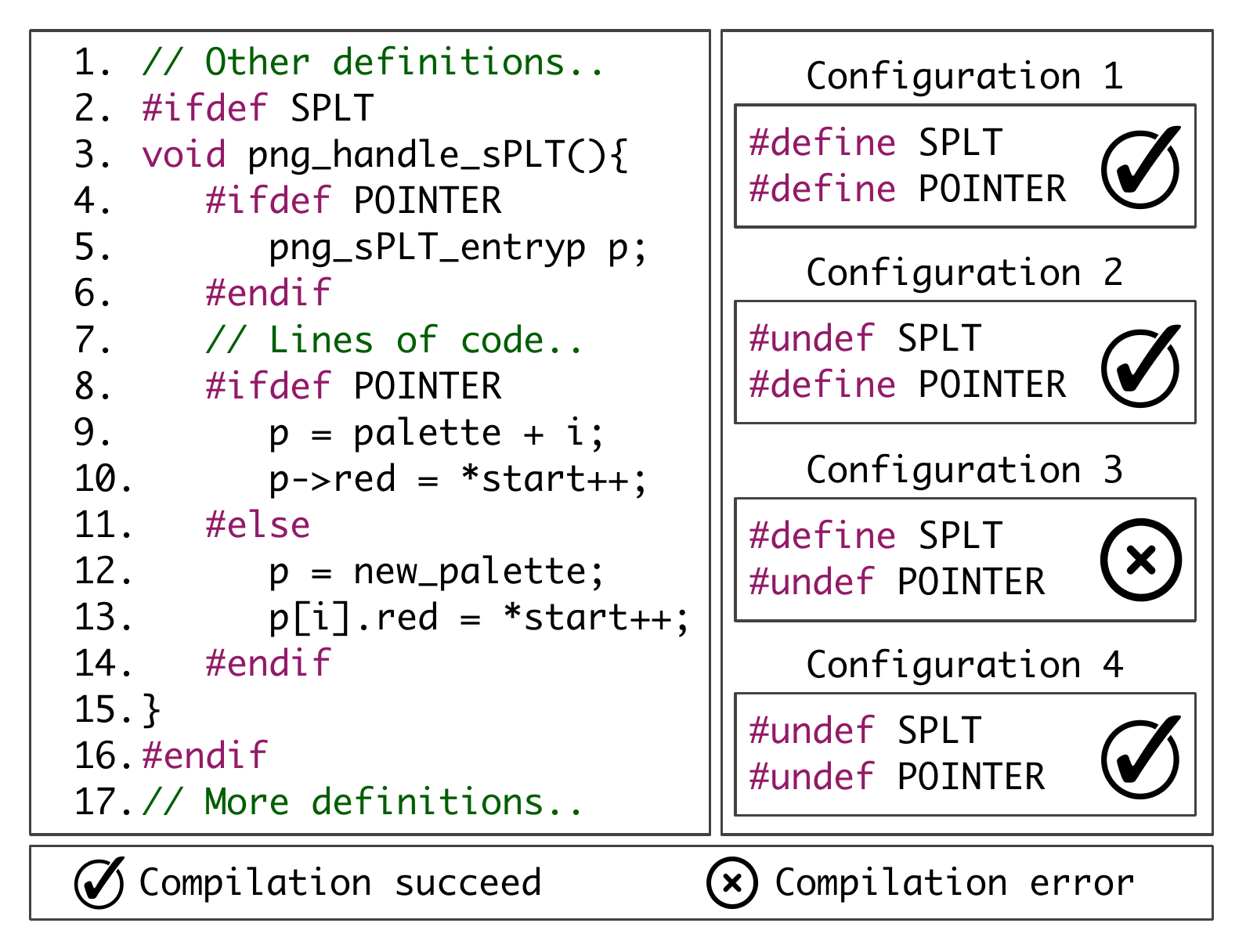}
\vspace{-0.2cm}
\caption{A fault in \textit{Libpng} that occurs when \texttt{SPLT} is enabled and \texttt{POINTER} is disabled.}
\label{fig:motivating-example}
\vspace{-0.3cm}
\end{figure}

Conditional compilation is used in many real-world systems to make the source code configurable~\cite{Liebig10}. 
For instance, Figure~\ref{fig:motivating-example} depicts a code snippet of \textit{Libpng}\footnote{\scriptsize \url{http://www.libpng.org/}} related to splitting images into segments. 
The splitting feature is optional and is included only when configuration option \texttt{SPLT} is enabled. 
This code snippet also contains a configuration option that checks for pointer-index support, controlled by configuration option \texttt{POINTER}. 
Using the C preprocessor, we can generate four different configurations from this code snippet: (1) both configuration options enabled; (2) only \texttt{POINTER} enabled; (3) only \texttt{SPLT} enabled; and (4) both configuration options disabled.

Most analysis tools for C code, such as \textit{gcc}, operate on preprocessed code, one configuration at a time.
By compiling the code snippet of Figure~\ref{fig:motivating-example} with \texttt{SPLT} enabled and \texttt{POINTER} disabled, we get a compilation error at Line 12. 
This line uses variable \texttt{p}, which is not declared before (Line 5) when we disable \texttt{POINTER}.
Because common analysis tools check only one configuration at a time, they do not show warning or error messages when one compiles the code depicted in Figure~\ref{fig:motivating-example} considering other configurations. 
This is an example of a configuration-related fault that can only be exposed in some combinations of configuration options~\cite{Yilmaz06,Kuhn04,Garvin11-2}. Unfortunately, the space of possible combinations is exponential, in the worst case, and it is usually too large to explore exhaustively.
For instance, the \textit{Linux Kernel} offers more than 12K configuration options, which give rise to more configurations than there are atoms in the universe.

To analyze real-world configurable systems, developers often use sampling algorithms that select only a few configurations for analysis.
For instance, one can check the code snippet presented in Figure~\ref{fig:motivating-example} using the \textit{most-enabled-disabled} sampling algorithm. It considers two configurations: (1) all configuration options enabled and (2) all options disabled.
However, it is not possible to detect the fault presented in Figure~\ref{fig:motivating-example} using the \textit{most-enabled-disabled} algorithm, as the fault requires enabling one configuration option while disabling another.
By using other sampling algorithms, one can detect this specific fault in \textit{Libpng}, but other faults possibly not.
For instance, one can use \textit{one-disabled}~\cite{Iago14}, which disables one configuration option at a time, or \textit{statement-coverage}~\cite{Tartler11-2}, which enables each block of optional code at least once.

Previous work~\cite{Garrido05,Garvin11-2,Iago14} has studied configuration-related faults similar to the one we discuss here and proposed many sampling algorithms~\cite{Iago14,Tartler11-2,Oster10,Perrouin10}.
However, researchers make assumptions that may not be realistic in practice.
For instance, they perform per-file instead of global analysis, and they ignore constraints between configuration options, header files and build-system information.
In this paper, we report on a comparative study of sampling algorithms initially accepting those assumptions (Section~\ref{s:comparative-study}), but explicitly evaluate the influence of including different types of information in a second study (Section~\ref{sec:limitations}).

%!TEX root = ../main.tex
\section{Study Design and Sampling\\Algorithms}
\label{sec:overallsetup}

Our overall goal is to compare state-of-the-art sampling algorithms regarding their capability to detect configuration-related faults and the size of their sample sets.
Furthermore, we study four assumptions of previous work, which often does not consider (1) constraints, (2) global analysis, (3) build-system information, and (4) header files. 
We perform our studies in the context of the C programming language and configuration options implemented with the C preprocessor (i.e., \texttt{\#ifdef}), as illustrated in the previous section.

%In particular, 
We aim at answering the following research questions:
\begin{compactitem}
    \item \textbf{RQ1.} What is the number of configuration-related faults detected by each sampling algorithm?
    \item \textbf{RQ2.} What is the size of the sample set selected by each sampling algorithm?
    \item \textbf{RQ3.} Which combinations of sampling algorithms maximize the number of faults detected and minimize the number of configurations selected?
   \item \textbf{RQ4.} What is the influence of the four assumptions on the feasibility to perform the analysis for each sampling algorithm?
    \item \textbf{RQ5.} What is the influence of the four assumptions on the number of faults detected by each sampling algorithm?
    \item \textbf{RQ6.} What is the influence of the four assumptions on the size of the sample set selected by each sampling algorithm?
\end{compactitem}

% We answer research questions \textbf{RQ1--3} with two studies of detecting configuration-related faults in a number of open source systems and subsequently answer questions \textbf{RQ4--6} with a third study analyzing the assumptions.

\subsection{Overall Study Design}
\label{sec:overalldesign}

At first glance, a study comparing sampling algorithms (RQ1--3) seems straightforward. We use a number of different sampling algorithms (independent variable) to measure how many of the faults we can find with them in different software systems and how big the sample set is (dependent variables). However, there are several challenges to overcome in the design of such an experiment.

Sampling the configuration space needs to be combined with a technique to detect faults in the respective selected configurations, such as inspection (unrealistically laborious), executing existing test suites (if available), automated test-case generation (looking for crashing defects), or static analysis (prone to false positives). 
If not conducted carefully, we might be evaluating the fault-detection technique instead of the sampling algorithm. 
We address this potential bias by taking the fault-detection technique out of the loop and by using a corpus of previously found configuration-related faults. For each known fault, we check whether the sampling algorithms select configurations in which the fault could have been found, assuming a suitable fault-detection technique. That is, by using a corpus of confirmed configuration-related faults, we eliminate the fault-detection technique as a confounding factor from our study setup.
However, we actually do not know if the sampling algorithms actually discovered more or different faults. 
We discuss this threat and an alternative study design in Section~\ref{sec:threats}.
 
%\item In Study 2, we use the static-analysis tool \emph{Cppcheck} as our fault-detection technique and measure how many warnings it reports that are configuration related. Although biased by the specific fault-detection technique, it complements our first study by providing a second perspective that does not rely on a manually curated corpus of faults.

%Both studies suffer from separate biases, but triangulating results across both studies helps us to gain confidence in the findings.

A second design challenge is how to evaluate the influence of the four assumptions (regarding global analysis, header files, constraints, and build-system information) behind many sampling algorithms. As we will show, lifting these assumptions can make it infeasible to apply some of the algorithms to real-world software systems. Therefore, we decided to proceed in two steps: First, we study tradeoffs among sampling algorithms (RQ1-3) under favorable conditions (i.e., fulfilling all assumptions). Subsequently, we investigate the influence of the assumptions (RQ4-6) on a smaller set of subject systems in a second study. The four assumptions are:
\begin{compactitem}

\item \textbf{Constraints:} Constraints between configuration options may exclude certain configurations (e.g., option X may only be selected if option Y is selected) from the set of valid configurations. 
A sample set may contain configurations that violate constraints. Unfortunately, configuration constraints are rarely documented explicitly---the \textit{Linux Kernel} is an exception and has been studied therefore extensively~\cite{Passos13,Tartler11}. 
In the presence of constraints, sample sets are often larger to achieve the same coverage, and highly optimized covering array tables\footnote{\scriptsize A covering array is a mathematical object used for software testing, which ensures specific coverage criteria. For example, a \textit{pair-wise} covering array ensures that all pairs of configuration options are considered by the array\cite{Yilmaz06,Johansen12}.} cannot be used. 
As we do not know configuration constraints for most of our subject systems, we exclude contraints entirely from the sampling process in our first study.

%\vspace{-0.1cm}

\item \textbf{Global analysis:} We can sample configurations per file or globally for the entire system. 
Even in systems with many configuration options, individual files are usually affected only by few options. 
Sampling over the global configuration space may detect inter-file faults (e.g., linker issues), but this often creates huge sample sets, which hardly affect individual files. 
Thus, in the first study, we assume a per-file analysis.

%\vspace{-0.1cm}

\item \textbf{Header files:} In C code, a significant amount of variability arises from header files. 
However, detecting all configuration options from header files in a sound way is a difficult and expensive task, which requires some form of variability-aware analysis~\cite{Adams07,Thum14,Kastner11,Dietrich12}.
It is necessary to resolve includes and macro expansions, but to keep the conditional directives (i.e., partial preprocessing).
We therefore analyze only configuration options inside source files in our first study. 

%\vspace{-0.1cm}

\item \textbf{Build system:} The build system may induce a significant amount of variability, such that certain files are not compiled in all configurations~\cite{Passos13,Dietrich12}. Since build systems are inherently difficult to analyze~\cite{NadiJournal2013}, we do not use build-system information in the first study.
\end{compactitem}

\subsection{Sampling Algorithms}
\begin{figure}[ht]
\centering
\vspace{-0.2cm}
\includegraphics[width=70mm]{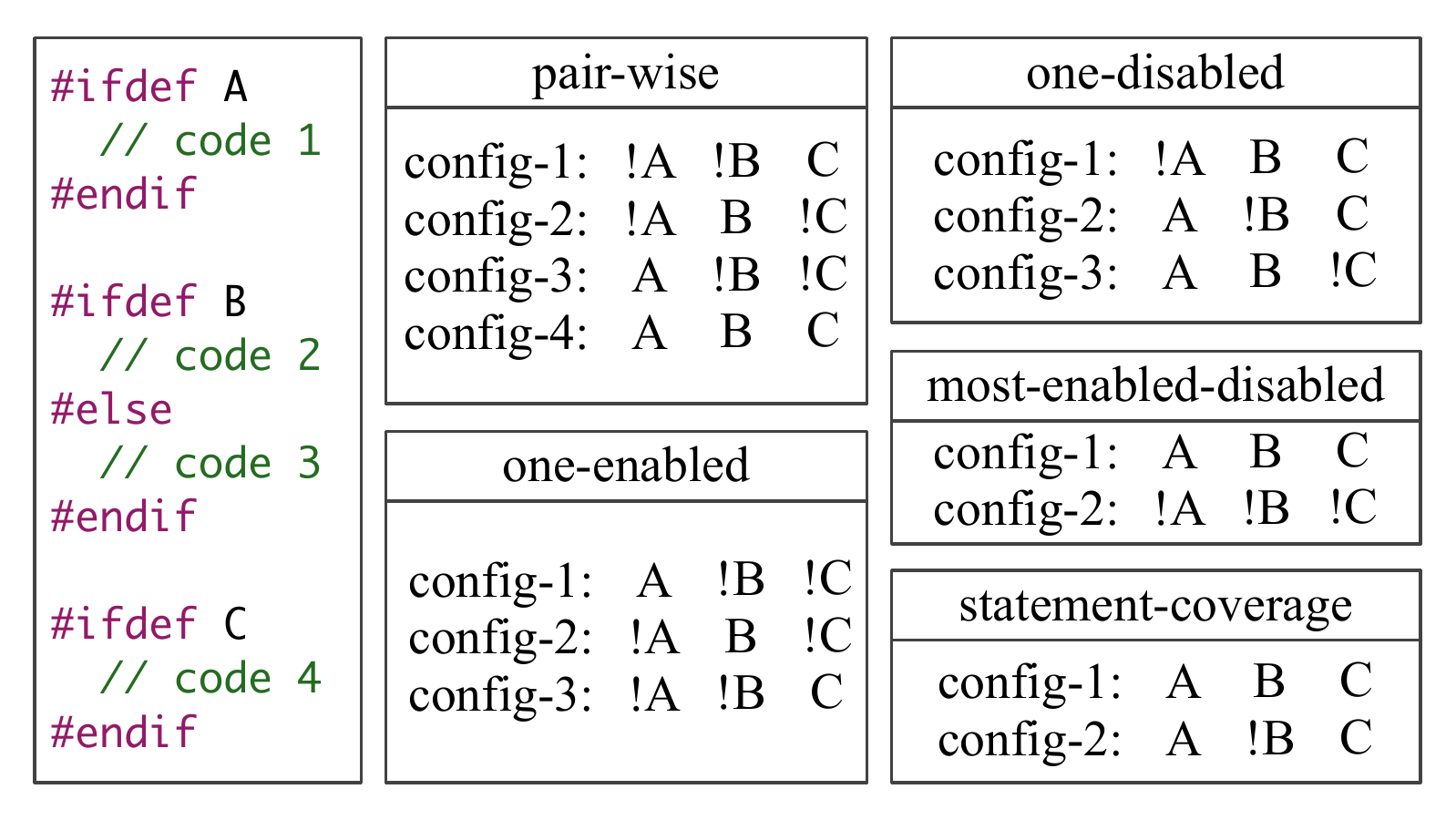}
\vspace{-0.2cm}
\caption{Comparing the sampling algorithms by example.}
\label{fig:sampling-explain}
\vspace{-0.5cm}
\end{figure}

In both studies, we will analyze the same set of \ALGORITHMS~sampling algorithms, proposed in prior work~\cite{Iago14,Tartler11-2,Oster10,Perrouin10,Johansen12,Marijan13,Liebig13} as well as their combinations.
We explain each sampling algorithm using the example code snippet of Figure~\ref{fig:sampling-explain}.

\begin{figure}[ht]
\centering
\vspace{-0.3cm}
\includegraphics[width=70mm]{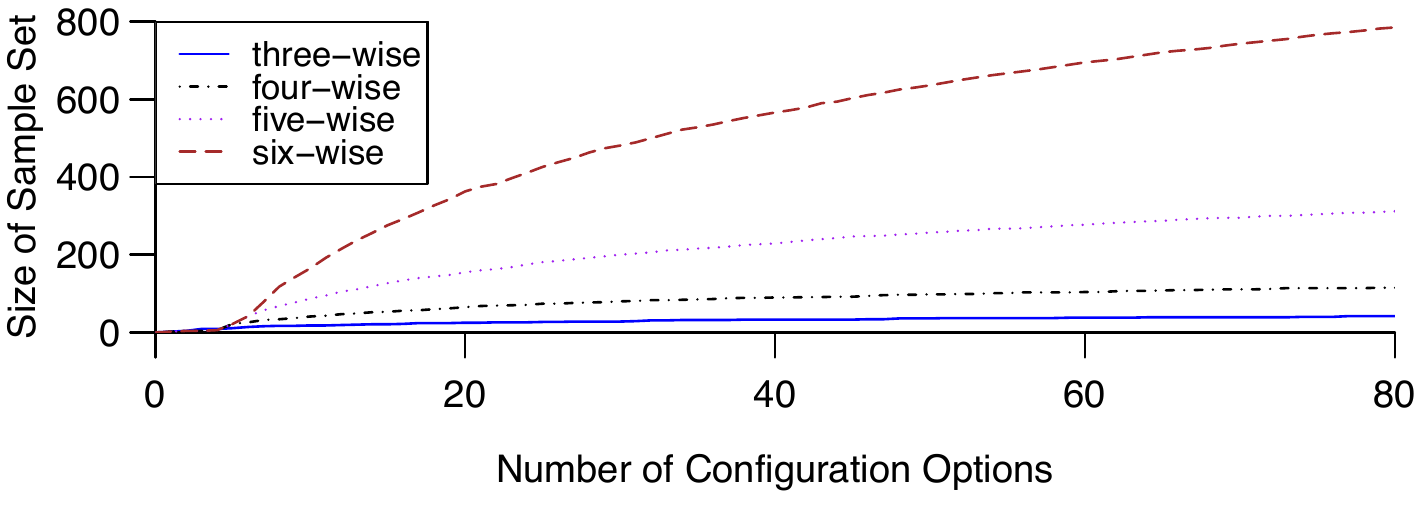}
\caption{Sample sets of \textit{t-wise} sampling considering a file with a number of configuration options ranging from zero to eighty.}
\label{fig:comparison-sampling-size}
\vspace{-0.3cm}
\end{figure}

\newpage
The \textbf{t-wise} algorithm covers all combinations of $t$ configuration options: \textit{pair-wise} checks all pairs of configuration options $(t=2)$~\cite{Oster10,Perrouin10,Johansen12,Marijan13}, and it selects four configurations of the example of Figure~\ref{fig:sampling-explain}. 
Considering options \textit{A} and \textit{B}, we can see that there is a configuration where both options are disabled (\textit{config-1}), two other configurations with only one of them enabled (\textit{config-2} and \textit{config-3}), and another configuration where both configuration options are enabled (\textit{config-4}). 
The same situation occurs for configuration options \textit{A} and \textit{C} and options \textit{B} and \textit{C}.
However, \textit{t} can take integer values to check different combinations of options, such as \textit{three-wise} $(t=3)$, \textit{four-wise} $(t=4)$, and \textit{five-wise} $(t=5)$. 
As we increase \textit{t}, the sizes of the sample sets also increase.
Figure~\ref{fig:comparison-sampling-size} presents the sample-set distributions of \textit{three-wise}, \textit{four-wise}, \textit{five-wise}, and \textit{six-wise}.
As we can see, \textit{three-wise} and \textit{four-wise} create small sample sets; \textit{five-wise} and \textit{six-wise} create much larger sample sets.
We selected samples based on precomputed and optimal covering-array tables\footnote{\scriptsize The precomputed and optimal covering arrays used in our study are available at \url{http://math.nist.gov/coveringarrays/}.} that select a minimal set of configurations that covers all \textit{t} combinations of configuration options.
These tables do not consider constraints between configuration options.
There are tools that implement \textit{t-wise} considering constraints, such as \textit{SPLCATool}~\cite{Johansen12}, \textit{CASA}~\cite{Garvin09}, and \textit{ACTS}~\cite{Borazjany12}. However, these tools do not necessarily select a minimal sample set or even guarantee \textit{t-wise} coverage, as discussed in Section~\ref{sec:limitations}.

The \textbf{statement-coverage} algorithm selects a set of configurations in which each block of optional code is enabled at least once~\cite{Tartler14}. We used \textit{statement-coverage} as implemented in the \textit{Undertaker}~\cite{Tartler11-2} tool suite.\footnote{\scriptsize Despite the existence of an algorithm to compute an optimal solution for the coverage problem~\cite{Liebig13}, which is \textit{NP-hard}, we used an algorithm that computes the sample set much faster, but may produce a sample set that is possibly larger than optimal.}
Notice that we are not using \textit{Undertaker} to detect dead code~\cite{Tartler11-2}, but to select configurations with the \textit{statement-coverage} algorithm.
As presented in Figure~\ref{fig:sampling-explain}, by enabling configuration options \textit{A}, \textit{B}, and \textit{C}, the algorithm ensures that the optional code blocks \texttt{code 1}, \texttt{code 2} and \texttt{code 4} are enabled at least once. 
However, it needs another configuration (e.g., \textit{A} and \textit{C} enabled, and \textit{B} disabled) to enable \texttt{code 3}.
Including each block of optional code at least once does not guarantee that all possible combinations of individual blocks of optional code are considered, though.

The \textbf{most-enabled-disabled} algorithm checks two samples independently of the number of configuration options. When there are no constraints among configuration options, it enables all options (\textit{config-1}), and then it disables all configuration options (\textit{config-2}).
\textbf{One-disabled} is an algorithm suggested by Abal et al.~\cite{Iago14} based on 42 faults found in the \textit{Linux Kernel}. It disables one configuration option at a time. We can also see in Figure~\ref{fig:sampling-explain} that it disables configuration option \textit{A} in \textit{config-1}, option \textit{B} in \textit{config-2}, and option \textit{C} in \textit{config-3}. In contrast, \textbf{one-enabled} enables one configuration option at a time. 
%Initially, we applied these algorithms without considering constraints, but we discuss the hurdles of considering constraints among configuration options in Section~\ref{sec:limitations}.

Finally, we implemented a \textit{random} sampling algorithm. 
Random sampling receives as input the maximum number of configurations (\textit{n}) to check per file. 
Then, it creates \textit{n} distinct configurations with all configuration options of the file and randomly assigns \texttt{true} or \texttt{false} for every option of each configuration.
For files which a \textit{brute-force} algorithm requires fewer configurations than the maximum number of configurations (\textit{n}) per file, random sampling selects all configurations.
We ran random sampling considering different numbers of configurations per file, ranging from 1 to 40. 
%Higher numbers require many configurations and would make random sampling inefficient.
For each number, we ran the analysis ten times and computed the average number of detected faults and the 95\% confidence interval.

\vspace{-0.2cm}
\section{Detecting Faults}
\label{s:comparative-study}

In this first study, we compared the fault-detection capabilities and the sample sizes of the~\ALGORITHMS~sampling algorithms using a corpus of \BUGSONE~known faults of 24 open-source systems to answer questions RQ1--3.
As explained in Section~\ref{sec:overallsetup}, we performed the first study under favorable assumptions, that is, without constraints, global analysis, build-system information, and header files. 

\begin{figure}[ht]
\centering
\vspace{-0.3cm}
\includegraphics[width=62mm]{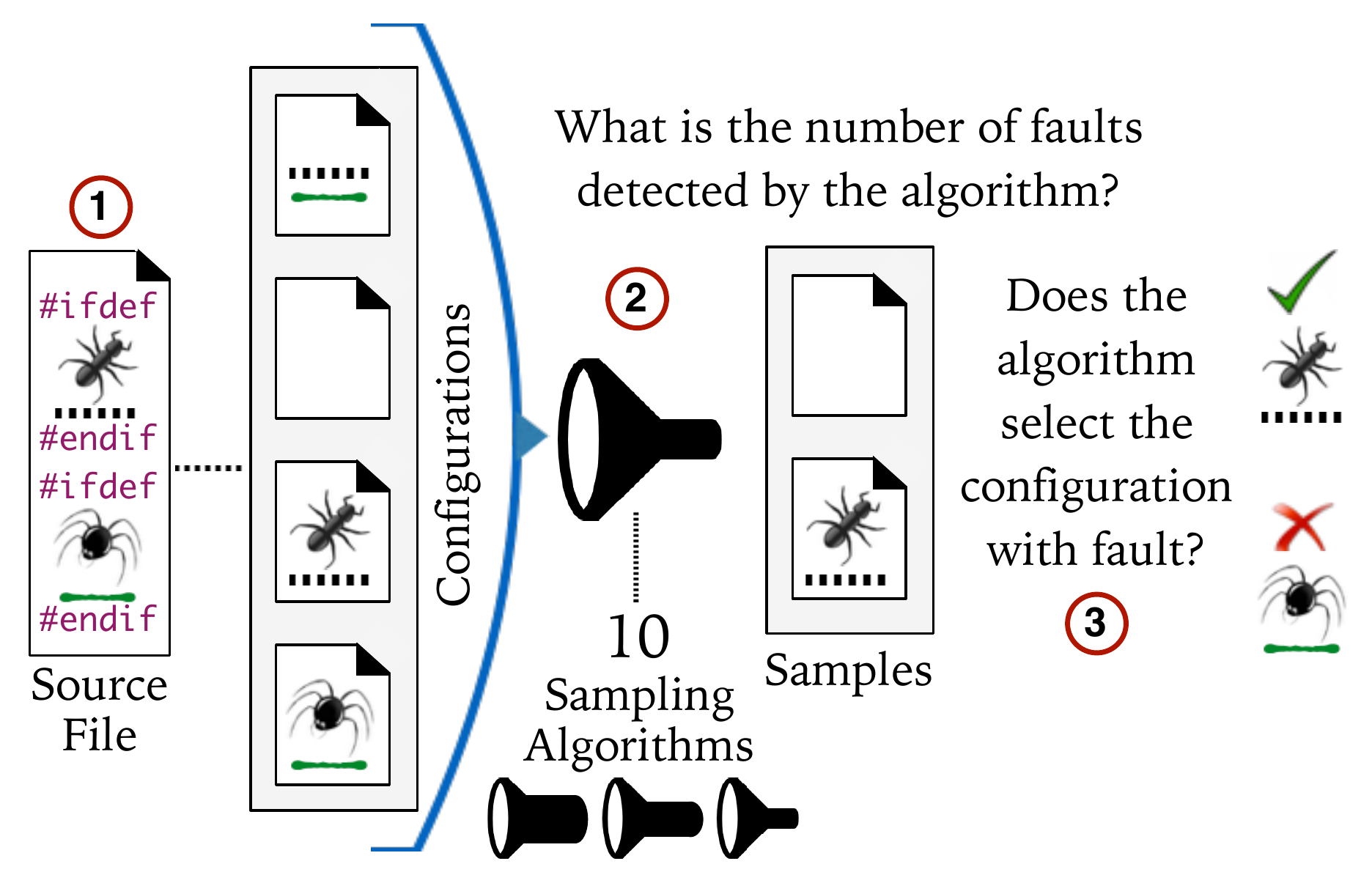}
\vspace{-0.2cm}
\caption{Strategy used to compare the sampling algorithms.}
\label{fig:comparison}
\vspace{-0.4cm}
\end{figure}

We proceeded in three steps, as illustrated in Figure~\ref{fig:comparison}.
In \textit{Step 1}, we select each source file of the given subject system. 
\textit{Step~2} applies each sampling algorithm to select the samples for every file.
\textit{Step~3} determines the number of configuration-related faults detected (RQ1) and measures the size of the sample set (RQ2) for each algorithm. The size of the sample set is the sum of the numbers of sampled configurations for every source file.
To identify the sampling algorithms that detect a fault, we consider its presence condition, which is a subset of system configurations in which the fault can be found~\cite{Alex15}, assuming a suitable fault-detection technique.
We checked whether we could find at least one configuration of this subset in the sampled configurations for each algorithm.
Finally, we repeat the process for combinations of sampling algorithms (RQ3).

\subsection{Corpus of Faults}

Using a corpus of configuration-related faults in a study raises the question of how to acquire a proper corpus and whether it is a representative corpus of configuration-related faults in real systems. Faults identified with existing sampling algorithms will obviously bias results toward these specific algorithms. Instead, we assembled a corpus of faults in which all faults have been identified in one of two ways:

\begin{compactitem}

\item Variability-aware analysis tools are able to identify certain kinds of faults (mostly syntax and type errors) by covering the entire configuration space without sampling. 
Difficulties in setting up these tools and narrow classes of detectable faults limit their applicability at this point, and their prototype status leads to false positives. 
We collected only configuration-related faults that have been reported by such tools, reported to the original developers, and confirmed or fixed by the developers~\cite{Kastner11,Medeiros13}.

\item We use configuration-related faults that have been manually identified and fixed by developers. 
Faults reported by users and fixed in the repository by the system's developers may be slightly biased toward more popular configurations, but are not systematically biased toward specific sampling algorithms. 
They represent configuration-related faults that are routinely detected and fixed in real software systems. We started with Abal's corpus of faults of the \textit{Linux Kernel}~\cite{Iago14}, and complemented it with faults found in other studies~\cite{Ribeiro14,Garvin11-2}, and our own investigation of software repositories (see Table~\ref{fig:faults}).

\end{compactitem}

\begin{table*}[t]
    \caption{Configuration-related faults considered in our first study.}
    \centering
    \begin{tabular}{ p{0.8cm} p{0.7cm} p{3.9cm} p{2.9cm} p{7.4cm} }
    \hline
    \small \textbf{Source} & \small  \textbf{Faults} & \small  \textbf{Kind} & \small  \textbf{Strategy} & \small  \textbf{Subject system (number of faults)}  \\ \hline
    \cite{Iago14} & 30 & \small Memory, type, and arithmetic & Repository mining & \small \textit{Linux} (30) \\ 
    \cite{Kastner11} & 10 & \small Syntax & \textit{TypeChef} & \small \textit{BusyBox} (10) \\ 
    \cite{Garvin11-2} & 5 & \small Include, and arithmetic & Repository mining & \small \textit{Gcc} (3), \textit{Firefox} (2) \\ 
    \cite{Ribeiro14} & 3 & \small Type & Repository mining & \small \textit{Gnome-keyring} (1), \textit{Gnome-vfs} (1), and \textit{Totem} (1) \\ 
     \cite{Medeiros13} & 22 & \small Syntax & \small \textit{TypeChef} & \small \textit{Apache} (3), \textit{Bash} (2), \textit{Dia} (2), \textit{Gnuplot} (5), \textit{Libpng} (3), and \textit{Libssh} (7) \\
    \hline
     \hspace{0.2cm}-  & 65 & \small Memory, type, and arithmetic & \small Our repository mining & \small \textit{Apache} (9), \textit{Bison} (2), \textit{Cherokee} (3), \textit{Cvs} (1), \textit{Dia} (1), \textit{Fvwm} (10), \textit{Gnuplot} (5), \textit{Irssi} (4), \textit{Libpng} (1), \textit{Lua} (1), \textit{Libssh} (10), \textit{Linux} (7), \textit{Libxml} (2), \textit{Lighttpd} (1),  \textit{Vim} (5), \textit{Xfig} (1), and \textit{Xterm} (2)\\
    \hline
    \small \textbf{Total} & 135 & \multicolumn{3}{c}{\hspace{-2.7cm}\small 70 faults collected from previous studies and 65 detected in our additional repository analysis.} \\
    \hline
    \end{tabular}
    \label{fig:faults}
    \vspace{-0.4cm}
\end{table*}

\begin{figure}[ht]
\vspace{-0.2cm}
\centering
\includegraphics[width=70mm]{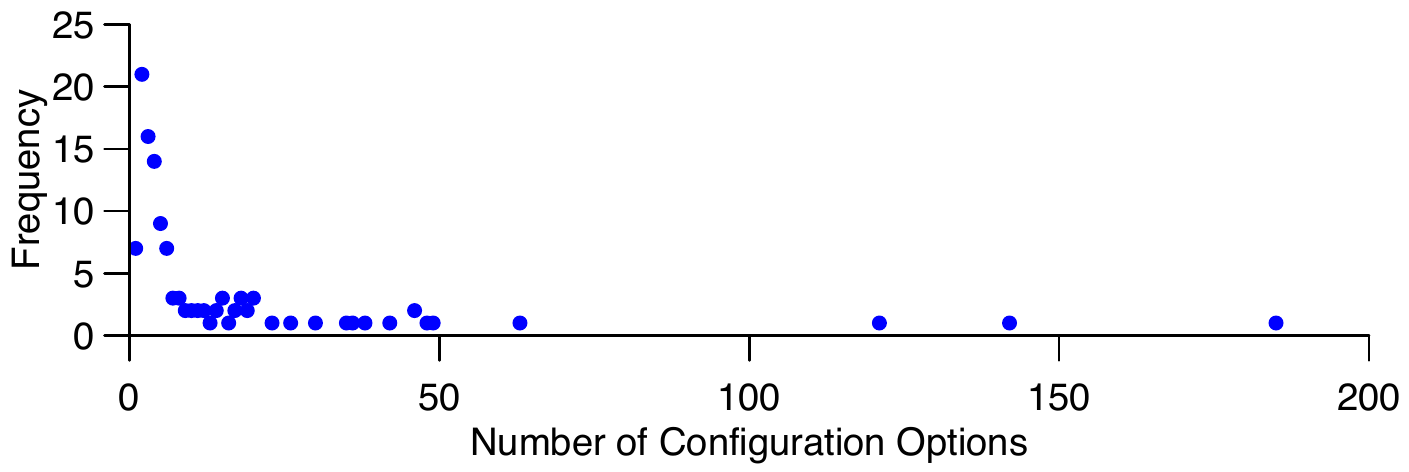}
\caption{Number of distinct configuration options in files with configuration-related faults.}
\label{fig:macros-files}
\vspace{-0.5cm}
\end{figure}

Overall, the corpus of faults used in our study includes \BUGSONE~configuration-related faults from 24 subject systems of various sizes and from different domains, over 125 different files with distinct numbers of configuration options (see Figure~\ref{fig:macros-files}).
Our corpus contains faults of different kinds, including 
syntax errors (34\%), 
memory leaks (22\%), 
null-pointer dereferences (17\%), 
uninitialized variables (13\%), 
undeclared variables and functions (5\%), 
resource leaks (3\%), 
array and buffer overflows (3\%), 
arithmetic faults (2\%), 
and type errors (1\%).
Table~\ref{subjects-table} presents a characterization of the subject systems we use in the first study, listing the project name, application domain, lines of code, number of files, number of configuration options, and number of known faults considered in our study.

Table~\ref{fig:conditions} shows the presence conditions of the faults and the number of configuration options that we need to enable or disable to detect the configuration-related faults we consider in the first study: for 78 faults (58\%), we need to enable some options; for 27 faults (20\%), we need to disable some configuration options; and for another 30 faults (22\%), we need to enable some options and disable others. 
The majority of faults (83\%) are related to one or two configuration options, while less than 5\% related to more than four configuration options. 

Notice that we discarded seven faults of the \textit{Linux Kernel} from our corpus that span multiple files, because we performed a per-file analysis in our first study.
We considered faults that require inter-procedural analysis, as long as all procedures are defined in the same file.

\subsection{Results and Discussion}
\label{sec-study1-results}
\begin{figure}[ht]
\centering
\vspace{-0.2cm}
\includegraphics[width=70mm]{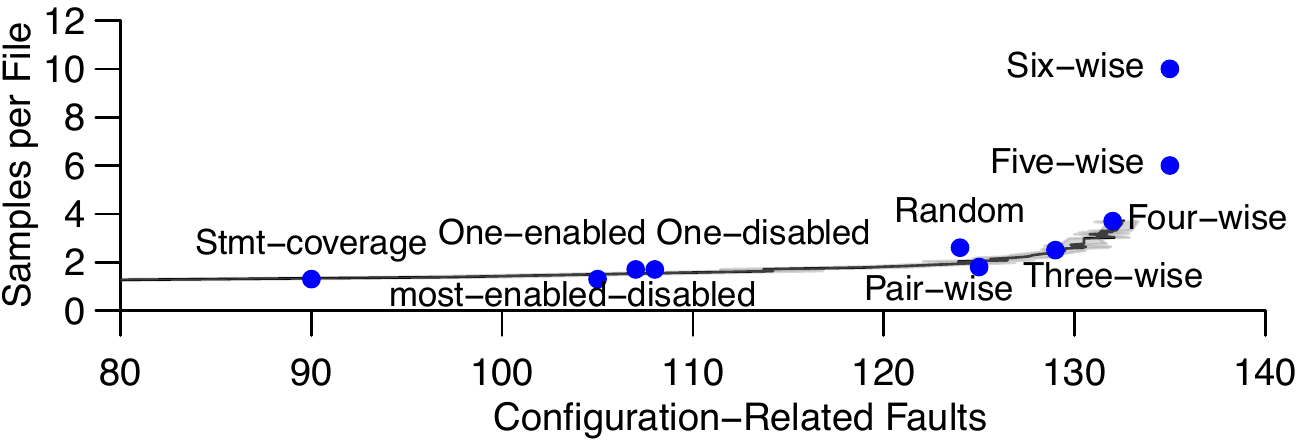}
\footnotesize{
\begin{tabular}{p{3.6cm} r r}
   \small \textbf{Sampling Algorithm} & \small \textbf{Faults} & \small \textbf{Samples/File} \\
    \hline
  \small Statement-coverage & \small 90 & \small 1.3 \\
  \small Most-enabled-disabled & \small 105 & \small 1.3 \\
  \small One-enabled & \small 107 & \small 1.7 \\
  \small One-disabled & \small 108 & \small 1.7 \\
  \small Random & \small 124 & \small 2.6 \\
  \small Pair-wise & \small 125 & \small 1.8 \\
  \small Three-wise & \small 129 & \small 2.5 \\
  \small Four-wise & \small 132 & \small 3.7 \\
  \small Five-wise & \small 135 & \small 6.0 \\
  \small Six-wise & \small 135 & \small 10.0 \\
  \hline
\end{tabular}
}
\caption{Number of configuration-related faults and samples per file for each sampling algorithm.}
\label{fig:sampling-analysis}
\vspace{-0.25cm}
\end{figure}

\begin{figure}[ht]
\centering
\includegraphics[width=74mm]{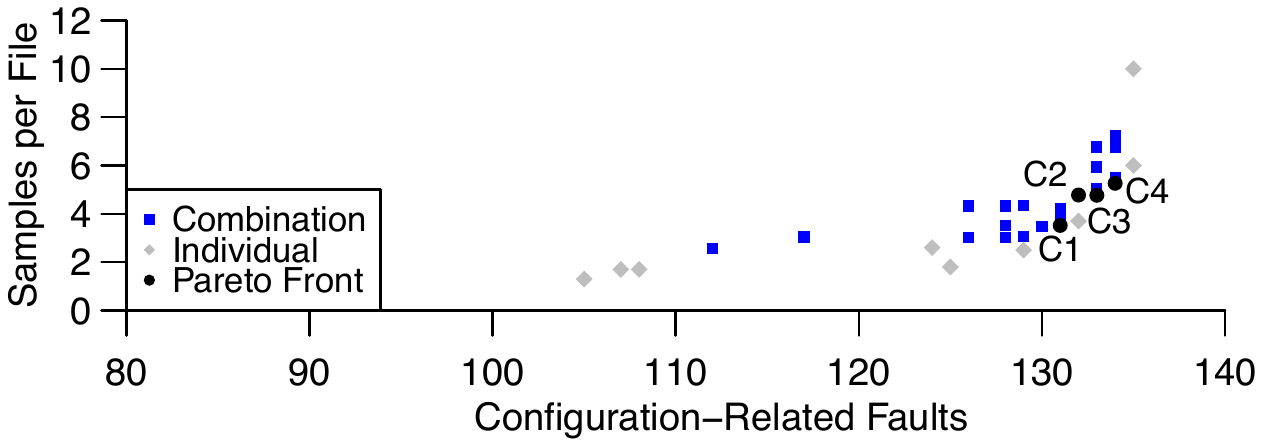}
\footnotesize{
\begin{tabular}{p{7.8cm}}
%\\
   \small \textbf{Sampling Algorithm} \\
    \hline
  \small C1 \small Pair-wise and one-disabled \\
  \small C2 \small One-enabled, one-disabled, and statement-coverage \\
  \small C3 \small One-enabled, one-disabled, and most-enabled-disabled \\
  \small C4 \small One-enabled, one-disabled, and pair-wise \\
  \hline
\end{tabular}
\begin{tabular}{p{0.1cm} r r p{0.1cm} r r}
\\
   \small \textbf{} & \small \textbf{Faults} & \small \textbf{Samples/File} & \small \textbf{} & \small \textbf{Faults} & \small \textbf{Samples/File} \\
    \hline
  \small C1 & \small 131 & \small 3.5 & \small C2 & \small 132 & \small 4.8\\
  \small C3 & \small 133 & \small 4.8 & \small C4 & \small 134 & \small 5.3\\
  \hline
\end{tabular}
}
\caption{Number of configuration-related faults and samples per file for the combination of algorithms on the \textit{Pareto Front}.}
\label{fig:combination}
\vspace{-0.4cm}
\end{figure}

For each sampling algorithm, we answered the research questions RQ1--2.
Figure~\ref{fig:sampling-analysis} presents the number of faults detected and the corresponding size of the sample set for each algorithm. Note that detecting more faults does not mean more efficiency, because there is a tradeoff between the number of faults detected and the size of the sample set.
We consider the efficiency of the sampling algorithms in terms of \textit{Efficiency}: $E = \textit{SizeOfSampleSet } / \textit{ NumberOfFaults}$. This ratio represents the number of configurations that one needs to check per fault to be detected. 
Furthermore, we analyzed 35 combinations of sampling algorithms to answer research question RQ3,
as illustrated in Figure~\ref{fig:combination}.  
We discuss the results in terms of the three research questions next.

\vspace{0.1cm}
\noindent
\textit{\textbf{RQ1.} What is the number of configuration-related faults detected by each sampling algorithm?} 
\vspace{0.1cm}

Overall, we found that all algorithms detected more than 66\% of all faults of our corpus.
\textit{Statement-coverage} detected the lowest number of faults, while \textit{six-wise} detected the highest number.
The majority of faults in our corpus can be detected by enabling or disabling fewer than six configuration options.
This way, \textit{six-wise} is able to detect all these faults. 
%There is one fault for which developers need to disable seven configuration options for triggering it, which \textit{six-wise} detected by chance. 
%Despite detecting the lowest number of faults, \textit{statement-coverage} checks fewer configurations than \textit{six-wise}.
\textit{Statement-coverage} missed 45 faults because they require developers to enable some configuration options and disable others (i.e., require specific combinations of multiple blocks of codes), whereas \textit{statement-coverage} is only concerned with including each block of code at least once in a system configuration.

All \textit{t-wise} sampling algorithms detected more than 92\% of the~\BUGSONE~configuration-related faults.
\textit{Six-wise} and \textit{five-wise} detected all faults.
\textit{Most-enabled-disabled}, \textit{one-enabled}, and \textit{one-disabled} detected all between 78\% to 80\% of the faults.
Furthermore, we present the average values of random sampling with a 95\% confidence interval (gray area) in Figure~\ref{fig:sampling-analysis}.
We ran random sampling with the maximum number of configurations per file (\textit{n}) ranging from 1 to 40, ten times for each value of \textit{n}.\footnote{\scriptsize Random selects 2.6 samples per file, on average.} 
We report the mean of all runs; it detected 124 (92\%) configuration-related faults.

%\newpage
\vspace{0.1cm} 
\noindent
\textit{\textbf{RQ2.} What is the size of the sample set selected by each sampling algorithm?} 
\vspace{0.1cm}

The sizes of the sample sets range from 1.3 to 10 configurations per file.
The algorithm \textit{most-enabled-disabled} selected the smallest sample set; \textit{six-wise} required the largest sample set (with more than 500K sampled configurations across all projects).
The number of configurations to check influences the time of analysis. So, it is not feasible to use algorithms with large sample sets in all situations, as we will discuss in Section~\ref{sec-guidance}. 
Based on our efficiency measure, we rank the algorithms starting from the most efficient: \textit{most-enabled-disabled},
\textit{pair-wise},
\textit{stmt-coverage},
\textit{one-disabled},
\textit{one-enabled},
\textit{three-wise},
\textit{random},
\textit{four-wise},
\textit{five-wise}, and
\textit{six-wise}. 
 
%For instance, the efficiency of the \textit{statement-coverage} algorithm is \textit{720}, while the efficiency of \textit{six-wise} is \textit{3,724}, even detecting all~\BUGSONE~faults, showing that \textit{statement-coverage} needs to check fewer configurations per fault detected (i.e., it is more efficient than \textit{six-wise}).

\vspace{0.1cm}
\noindent
\textit{\textbf{RQ3.} Which combinations of sampling algorithms maximize the number of faults detected and minimize the number of configurations selected?}
\vspace{0.1cm}

In addition to the individual algorithms, we analyzed combinations (that is, the union of the sample sets produced by the respective sampling algorithms) of two and three sampling
algorithms, excluding \textit{random}, \textit{five-wise}, and \textit{six-wise} algorithms.
We excluded \textit{random} because it detects different numbers of faults in different runs, and we excluded \textit{five-wise} and \textit{six-wise} because they already detected all~\BUGSONE~faults.
Furthermore, we excluded combinations with more than three algorithms, because they resulted in inefficient combinations according to our efficiency function.

% Table 2 - Subject Characterization and Number of Syntax Errors in Releases
\begin{table*}[t]
\caption{Project characterization and the total number of known faults used in the first study.}
\centering
\begin{tabular}{p{2.5cm} p{4.5cm} r r r r r r}
\hline

\textbf{Project}  &  \textbf{Application domain}  &  \textbf{LOC} &  \textbf{Files} & \textbf{Configuration options} & \textbf{Faults} \\
\hline    
\textit{Apache}  &  \textit{Web} server  &   144,768   &  362  & 700 &   12 \\
\textit{Bash}  &  language interpreter  &   44,824   &  138 & 1,427 & 2  \\
\textit{Bison}  &  parser generator  &   24,325   &  129  & 269 &  2 \\
\textit{Busybox} & UNIX utilities & 189,722 & 805 & 1,418 & 10 \\
\textit{Cherokee}  &  \textit{Web} server  &   63,109   &  346 & 452 &  3  \\
\textit{Cvs} & version control system & 76,125 & 236 & 628 & 1 \\
\textit{Dia}  &  diagramming software  &   28,074   &  132 & 307 &  3 \\
\textit{Firefox} & \textit{Web} browser & 6,017,673 & 22,423 & 17,415 & 2 \\
\textit{Fvwm}  &  windows manager  &   102,301   &  270  & 301 &  10 \\
\textit{Gcc} & C/C++ compiler & 1,946,622 & 22,034 & 3,825 & 3 \\
\textit{Gnome-keyring} & daemon application  & 76,525 & 376 & 213 & 1\\
\textit{Gnome-vfs} & file system library & 78,380 & 286 & 427 & 1 \\ 
\textit{Gnuplot}  &  plotting tool  &   79,557   &  152  & 500 &  10 \\ 
\textit{Irssi}  &  IRC client  &   51,356   &  308  & 157 &  4  \\
\textit{Libpng}  &  PNG library  &   44,828   &  61  &  327 &  4 \\
\textit{Libssh}  &  SSH library  &   28,015   &  125  & 115 &   17 \\
\textit{Libxml}  &  XML library  &   234,934   &  162 &  2,126 &  2 \\
\textit{Lighttpd}  &  \textit{Web} server  &   38,847   &  132  & 215 &  1 \\
\textit{Linux} & operating system & 12,594,584 & 37,520 & 26,427 & 37 \\
\textit{Lua}  &  language interpreter &   14,503   &  59 & 145 & 1 \\ 
\textit{Totem} & movie player & 31,596 & 135 & 84 & 1 \\
\textit{Vim}  &  text editor  &   288,654   &  178  &  942 &  5 \\
\textit{Xfig} & vector graphics editor & 70,493 & 192 & 143 & 1 \\
\textit{Xterm} & terminal emulator & 50,830 & 58 & 501 & 2 \\
\hline
 \textbf{Total} &   &   &    &  &  \textbf{135}\\ 

%\multicolumn{6}{c}{{Table labels here!}} \\
\hline
\end{tabular}
\label{subjects-table}
\vspace{-0.5cm}
\end{table*}

Figure~\ref{fig:combination} relates the number of faults and the size of sample sets for all combinations of sampling algorithms.
Based on the results, we determined the \textit{Pareto Front} to illustrate tradeoffs between number of detected faults and size of the sample sets.
Figure~\ref{fig:combination} also presents the combinations of sampling algorithms on the \textit{Pareto Front}, starting from the most efficient: \textit{C1}, \textit{C3}, \textit{C2}, and \textit{C4}.
%According to our efficiency function, we rank these four combinations starting from the most efficient: C1, C3, C2, and C4.
%On the other hand, the combination of \textit{one-enabled}, \textit{one-disabled} and \textit{four-wise} is the least efficient.

%\vspace{-0.2cm}
\summary{All sampling algorithms are able to detect at least 66\% of the configuration-related faults; \textit{most-enabled-disabled}, \textit{pair-wise}, and \textit{statement-coverage} are the most efficient algorithms; some combinations  provide a useful balance between sample size and fault-detection capabilities.}
%\vspace{-0.3cm}

%!TEX root = ../main.tex
\vspace{-0.3cm}
\section{Effects of Assumptions}
\label{sec:limitations}

In the first study, we made many simplifying assumptions also made in related studies on sampling~\cite{Shi05,Kuhn04,Lei08}.
We ignored constraints, header files, and build-system information, and we did a per-file analysis only.
In more realistic conditions, these assumptions often do not hold: For example, 
constraints often exist, and ignoring them may lead to false positives, but constraints are rarely documented systematically and therefore easily ignored. Similarly 
information from build systems may increase precision but build systems are inherently difficult to analyze~\cite{Adams07,Dietrich12}. 
While the simplifying assumptions allow researchers and practitioners to apply sampling algorithms quickly to a large set of systems, as we did in our first study, their influence on practicability and effectiveness is not well understood. 
Therefore, in a second study, we explore the effect of each assumption on the efficiency of the sampling algorithms.

Basically, we replicate the first study for a subset of the corpus, investigating how the assumptions affect each sampling algorithm (\textbf{RQ4--6}).
To increase internal validity~\cite{Siegmund15}, we considered each assumption separately as an independent variable that we manipulate to understand the influence of each assumption on sampling.
We limit the second study to faults of the \textit{Linux Kernel} and \textit{BusyBox} (47 faults from the first study), because these subject systems are the only ones for which we have build-system and constraint information from the \textit{LVAT} and \textit{TypeChef} projects~\cite{She10,Berger10,NadiTSE15}. For the \textit{Linux Kernel}, we consider additionally seven known faults that cross files, which we excluded from our original corpus, as we discussed in Section~\ref{s:comparative-study}.

Table~\ref{fig:summary-dimensions} summarizes the number of configuration-related faults detected, sizes of sample sets, and the ranking of sampling algorithms per lifted assumption.

\subsection{Constraints}
\label{sec:constraints}

Constraints exclude certain combinations of configuration options (e.g., option X must be selected if option Y is selected) from the set of valid configurations. Faults identified in invalid configurations are considered false positives (which did not occur in the first study, because we consider only a corpus of true positives); hence sampling invalid configurations adds no value.
The analyzed version of the \textit{Linux Kernel} has 293,826 constraint clauses among its configuration options; \textit{BusyBox} has 615.

%\TODO{if you take the original sample set from Study 1 and discard all configurations in there that are valid, how many samples would remain? how many bugs would they find?}

In the original sample sets of the first study, many sampled configurations are actually 
invalid in these highly constrained configuration spaces.
For instance, \textit{random} selects 24\% of valid configurations and the percentage goes up to 43\% when picking \textit{most-enabled-disabled}.
Sampling within such constrained spaces is more challenging for all sampling
algorithms, as solvers or search-based strategies are needed. We incorporate constraints as follows:
\begin{compactitem}

    \item \textit{Most-enabled-disabled:} We cannot simply enable all options if some of them are mutually exclusive. 
    Instead, we use a solver to find two valid configurations with the maximum number of configuration options enabled and disabled.
    If there are multiple optimal solutions, we pick the first offered by the solver.

    \item \textit{One-enabled/disabled:} Similarly, for each option, we use a solver to identify the 
    valid configuration that disables/enables the most other options.

    \item \textit{Random sampling:} 
    We randomly assigned \textit{true} or \textit{false} for every configuration option inside a file
    and discard invalid assignments until we find the desired number of configurations.
    Truly random sampling in large constrained spaces with many options is still 
    a research problem though, with recent progress in theory~\cite{Chakraborty15} and recent pragmatic search heuristics~\cite{Henard15}. 

    \item \textit{Statement-coverage:} To select a minimal set of covering configurations, we need to consider constraints. Conceptually we can use the original implementation of \textit{statement-coverage}, as part of \textit{Undertaker}~\cite{Tartler11-2}, but the tool is not flexible to handle other projects than \textit{Linux}. Thus, we use an alternative implementation that we created in previous work~\cite{Liebig13}.

    \item \textit{T-wise sampling:} The covering array tables used in the first study are precomputed,
    often optimal solutions that, however, assume independence of all options.
    Recent research investigated strategies to generate \textit{t-wise} covering arrays 
    for constrained configuration spaces, such as \textit{SPLCATool}~\cite{Johansen12}, \textit{CASA}~\cite{Garvin09}, and \textit{ACTS}~\cite{Borazjany12}.
    All tools use heuristics and may produce larger-than-optimal sampling sets and
    sampling sets that do not actually achieve full t-wise coverage.
    To generate the \textit{pair-wise} covering array, we used \textit{SPLCATool}.
    %, but we found pairs of configuration options that are not covered in the array, even though they were not excluded by constraints.
    We failed to generate \textit{three-wise} or even higher covering arrays for the \textit{Linux Kernel}: 
    Even with 120 Gb RAM we ran out of memory; a developer from \textit{CASA} estimated that that the generation 
    could take months and would require a 1.6 Tb array to track the covered options.
    Overall, we could not find an alternative to implement the \textit{three-wise}, \textit{four-wise}, \textit{five-wise} and \textit{six-wise} algorithms considering constraints; existing approaches are intractable for the size and complexity of the \textit{Linux Kernel}.
\end{compactitem}
%Depending on the solver or search strategy, the identified samples may not be optimal. 

The changes in sampling algorithms to incorporate constraints changed the efficiency of
the algorithms as summarized in Table~\ref{fig:summary-dimensions}.
Most affected were t-wise strategies:
\textit{Pair-wise} required a larger sample set and detected fewer faults (including faults that \textit{pair-wise} should have guarantee to find)
from the \textit{Linux Kernel}, because the used heuristics are  
unsound and do not cover all valid pairs of options.
\textit{Three-wise} sampling and beyond was not tractable at all.

The time to compute sample sets increases significantly when adding constraints. 
Our use of a SAT solver required significant additional time and memory to generate the sample sets.
On average, we created sample sets for each file in 0.04 seconds without constraints (the first study), while the analysis with constraints took 0.75 seconds per file, on average.
This time represents an increase from 15~minutes to over 4~hours for the \textit{Linux Kernel}.
Regarding the ranking of algorithms, \textit{most-enabled-disabled} and \textit{statement-coverage} remain at top positions (see Table~\ref{fig:summary-dimensions}); the \textit{t-wise} algorithms dropped significantly or were not feasible at all.

%\TODO{summary:} essential reduction of false positives; high costs for generating sample sets,
%often not optimal or unique;
%infeasible for three-wise and higher.

\vspace{-0.1cm}

\summary{When considering constraints, we substantially reduce false positives; but there are high costs for generating sample sets, which are often not optimal.}

\vspace{-0.1cm}
 
\begin{table}
    \caption{Presence conditions of the configuration-related faults.}
  
    \centering
    \begin{tabular}{ p{6cm} >{\centering\arraybackslash}p{1.4cm} }
    \hline
    Some configuration options enabled & 78 (58\%) \\ \hline
    \small{$a$} & 59 \\
    \small{$a$ $\wedge$ $b$} & 13 \\
    \small{$a$ $\wedge$ $b$ $\wedge$ $c$} & 5 \\
    \small{$a$ $\wedge$ $b$ $\wedge$ $c$ $\wedge$ $d$ $\wedge$ $e$} & 1 \\
    \hline
    Some configuration options disabled & 27 (20\%) \\ \hline
    \small{$!a$} & 16 \\
    \small{$!a$ $\wedge$ $!b$} & 8 \\
    \small{$!a$ $\wedge$ $!b$ $\wedge$ $!c$} & 1 \\
    \small{$!a$ $\wedge$ $!b$ $\wedge$ $!c$ $\wedge$ $!d$} & 1 \\
    \small{$!a$ $\wedge$ $!b$ $\wedge$ $!c$ $\wedge$ $!d$ $\wedge$ $!e$ $\wedge$ $!\text{f}$ $\wedge$ $!g$} & 1 \\
    \hline
    Some options enabled and some disabled\hspace{0.5cm} & 30 (22\%) \\ \hline
   \small{($!a$ $\wedge$ $b$) $\vee$ ($a$ $\wedge$ $!b$)} & 17 \\
    \small{($a$ $\wedge$ $b$ $\wedge$ $!c$) $\vee$ ($!a$ $\wedge$ $!b$ $\wedge$ $c$)} & 6 \\
    \small{($a$ $\wedge$ $b$ $\wedge$ $!c$ $\wedge$ $!d$) $\vee$ ($a$ $\wedge$ $b$ $\wedge$ $c$ $\wedge$ $!d$)} & 3 \\
    \small{($a$ $\wedge$ $b$ $\wedge$ $c$ $\wedge$ $d$ $\wedge$ $!e$) $\vee$ ($!a$ $\wedge$ $!b$ $\wedge$ $!c$ $\wedge$ $!d$ $\wedge$ $e$)} & 2 \\
    \small{$a$ $\wedge$ $!b$ $\wedge$ $!c$ $\wedge$ $!d$ $\wedge$ $!e$ $\wedge$ $!\text{f}$} & 1 \\
    \small{$a$ $\wedge$ $b$ $\wedge$ $!c$ $\wedge$ $!d$ $\wedge$ $!e$ $\wedge$ $!\text{f}$} & 1 \\
    \hline
   \multicolumn{2}{c}{\vspace{-0.1cm} \includegraphics[width=80mm]{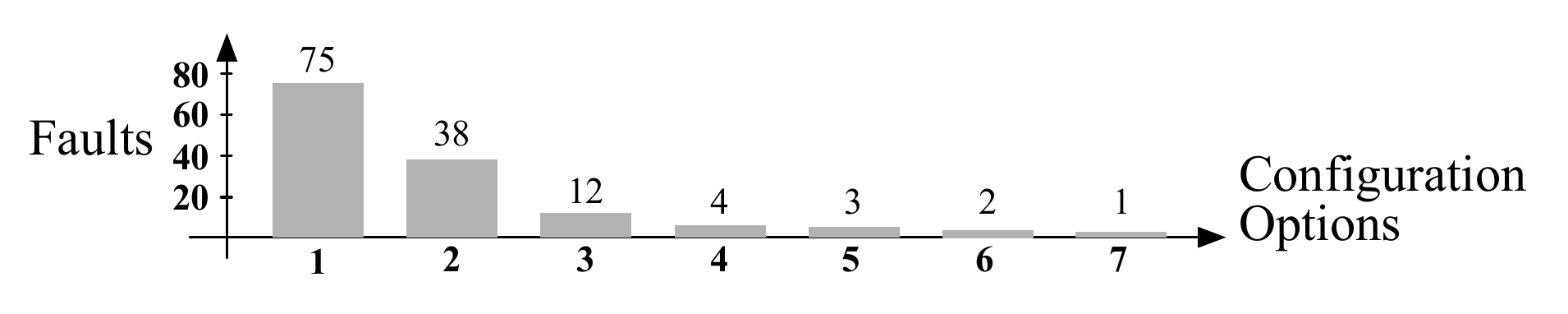}} \vspace{-0.1cm} \\
    \hline
    \end{tabular}
    \label{fig:conditions}
	\vspace{-0.6cm}
\end{table}  

%\vspace{-0.6cm}
\subsection{Global Analysis}
\label{sec:global}

To perform global analysis, we created a single sample set across all files,
instead of a distinct set per file. Such global set allows us to perform
cross-file analysis to find faults that cannot be identified on a per-file basis, such
as linking problems. However, for global analysis, a sampling algorithm needs
to consider all options in the system, not just the subset of options used in
each file.

We were not able to generate global sample sets with any \textit{t-wise} algorithm
at the scale of our subject systems. The largest precomputed tables we found covered
up to 2K options (pair-wise) or 191 options (six-wise). 
We are not aware of any tool that has the capability to generate covering arrays 
for such a large number of configuration options, even without constraints.
\textit{Statement-coverage} also turns intractable, as it requires to solve the coverage problem considering all source files of the project (i.e., equivalent to concatenating all source code into a single file and finding a set of configurations that enabled all optional code blocks at least once).
\textit{One-enabled} and \textit{one-disabled} require substantially larger
sample sets, as more options are considered
(from 1.7 to almost 8K). \textit{Random} requires
larger sample sets, on average, because previously we could use smaller
sample sets when the file had only few options.
\textit{Most-enabled-disabled} is the only algorithm for which the size of sample sets was not influenced, because it is not sensitive to the number of options and it always selects exactly two configurations.

\begin{table*}
	\small
    \caption{Number of faults, size of sample sets and ranking considering the 47 faults of the second study.}
    \centering
    \begin{tabular}{p{3.2cm}|p{0.6cm}p{0.8cm}p{0.6cm}|p{0.6cm}p{1cm}p{0.6cm}|p{0.6cm}p{1cm}p{0.6cm}|p{0.6cm}p{0.8cm}p{0.6cm}}
    \hline
\textbf{Algorithms} & \multicolumn{3}{c|}{\textbf{Constraints}} & \multicolumn{3}{c|}{\textbf{Global analysis}} & \multicolumn{3}{c|}{\textbf{Header files}} & \multicolumn{3}{c}{\textbf{Build system}} \\
\hline
 & Faults & Configs &  Rank & Faults & Configs & Rank & Faults & Configs &  Rank & Faults & Configs & Rank \\
Pair-wise &  
				33 \scriptsize{$\downarrow$} & 30 \scriptsize{$\Uparrow$} & 5 & 
				-- & -- & -- &
				39 $=$ & 936 \scriptsize{$\Uparrow$} & 4 & 
				33 \scriptsize{$\downarrow$} & 2.8 \scriptsize{$\uparrow$} & 4\\
Three-wise & 
				  -- & -- &  -- & 
				   -- & -- & -- &
				  43 $=$ & 1,218 \scriptsize{$\Uparrow$} & 5 &
				   42 \scriptsize{$\downarrow$} & 3.9 \scriptsize{$\uparrow$} & 5\\
Four-wise & 
			 -- & -- &  -- & 
			   -- & -- & -- &
			   45 $=$ & 1,639 \scriptsize{$\Uparrow$} & 7 &
			   45 $=$ & 5.7 \scriptsize{$\uparrow$} & 8\\
Five-wise & 
				 -- & -- &  -- & 
				 -- & -- & -- & 
				 -- & -- & -- &
				 47 $=$ & 8.3 \scriptsize{$\uparrow$} & 9\\
Six-wise & 
				 -- & -- &  -- & 
				 -- & -- & -- & 
				 -- & -- & -- &
				 47 $=$ & 12 \scriptsize{$\uparrow$} & 10 \\
Most-enabled-disabled &  
			23 \scriptsize{$\downarrow$} & 1.4 $=$ &  1 &  
			27 $=$ & 1.4 $=$ &  1 &
			27 $=$ & 1.4 $=$ &  1 &
			26 \scriptsize{$\downarrow$} & 1.4 \scriptsize{$\uparrow$} &  2\\
One-enabled &  
				30 \scriptsize{$\uparrow$} & 1.1 \scriptsize{$\downarrow$}  & 3 & 
				31 \scriptsize{$\uparrow$} & 7,943 \scriptsize{$\Uparrow$}  & 3 &
				31 \scriptsize{$\uparrow$} & 890 \scriptsize{$\Uparrow$} & 6 &
				20 \scriptsize{$\downarrow$} & 2.3 \scriptsize{$\uparrow$} &  7\\
One-disabled &  
				38 \scriptsize{$\downarrow$} & 1.1 \scriptsize{$\downarrow$} &  4 &  
					39 $=$ & 7,943 \scriptsize{$\Uparrow$}  & 2 &
					39 $=$ & 890 \scriptsize{$\Uparrow$} &  3 &
					39 $=$ & 2.3 \scriptsize{$\uparrow$} &  3\\
Random &  
			39 \scriptsize{$\downarrow$}  & 4.1 $=$ &  6 &  
				29 \scriptsize{$\Downarrow$} & 8,123  \scriptsize{$\Uparrow$} &  4 &
				40 \scriptsize{$\downarrow$} & 17.2  \scriptsize{$\Uparrow$} &  2 &
				41 $=$ & 4.2  \scriptsize{$\uparrow$} &  6\\
Stmt-coverage &  
				32 \scriptsize{$\uparrow$} & 4.1 \scriptsize{$\uparrow$} &  2 & 
				-- & -- & -- & 
				-- & -- & -- &
				25 $=$ & 1.3 \scriptsize{$\uparrow$} &  1 \\
   \hline    
    \multicolumn{13}{c}{\scriptsize{Some algorithms do not scale, indicated using dashes (--). We use $\uparrow$ and $\downarrow$ to represent small changes in the number}} \\ 
    \multicolumn{13}{c}{\scriptsize{of faults and size of sample set, as compared to our first study and we use $\Uparrow$ and $\Downarrow$ to represent larger changes.}} \\ 
    %\TODO{I'd suggest to have two different kinds of arrows for small and large changes (e.g. filled or not/ single or double)}
    \hline
    \end{tabular}
\label{fig:summary-dimensions}
\vspace{-0.4cm}
\end{table*}

To explore the ability of global analysis to identify cross-file faults,
we opportunistically analyzed 7~known faults of the \textit{Linux Kernel}~\cite{Iago14} that span multiple files,
which we had to exclude from our first study.
We detected all seven faults by applying \textit{one-enabled} and \textit{one-disabled} with global analysis.
\textit{Most-enabled-disabled} detected five (71\%) out of the seven faults, and \textit{random} detected four (57\%) faults.
The other algorithms are not feasible with global analysis.

%\TODO{summary:} potential to detect non-modular faults that span multiple files; explosion in the number of considered options leads to large sample sets; too large for n-way and stmt-coverage.

\summary{Using a global analysis, we can potentially detect faults that span multiple files; it causes an explosion in the number of configuration options that leads to large sample sets, too large for \textit{t-wise} and \textit{statement-coverage}.}

\subsection{Header Files}
\label{sec:hf}

In C source code, variability may be introduced by header files, because macros used in \texttt{\#ifdefs} can have non-local effect. 
If sampling is applied only to variability in the main C source file, faults stemming from variability in header files
may not be detected. For example, a function may not be declared in all configurations of the header, a type name may 
be defined as either \texttt{int} or \texttt{long} depending on configuration decisions in the header, or
a macro may be defined in the header only in some configurations.
Precisely analyzing header variability is challenging, though, due to the interaction of \texttt{\#include} directives with
conditional compilation and macros. Precise analyses exist~\cite{Kastner11,Gazzillo12}, but are challenging
and time-consuming to use, because one needs to set up the environment with all header files used by the project.

Incorporating header files increases the number of configuration options per file significantly. 
For instance, whereas the files of the \textit{Linux Kernel} contain, on average, 3~distinct configuration options
when ignoring variability from header files, headers add another 238 distinct configuration options, on average. 
This increases the size of the sample set for all algorithms, except for \emph{most-enabled-disabled}.
For \textit{statement-coverage}, \textit{five-wise}, and \textit{six-wise}, our subject systems reach configuration spaces
for which these algorithms become intractable.

%The other algorithms are feasible when including header files.
Since our corpus does not include faults caused by misconfigurations from header files,
most algorithms detect the same faults.
The \textit{one-enabled} algorithm detected more faults, because including configuration options from headers allowed it to disable more options, while enabling one at a time.

%\TODO{summary:} potential to detect additional faults from header files; difficult setup; much larger sample sets (if feasible at all) lead to ranking changes.

\summary{When incorporating header files, there is a potential to detect additional faults from header files; but the setup is difficult and the sample sets are much larger (if feasible at all), which lead to ranking changes.}

%The \textit{t-wise} algorithms are influenced negatively, because they require larger covering array tables with more configurations per file, which decreases their efficiency.
%Conversely, \textit{one-disabled} got a better ranking position because of the points discussed previously.
%\textit{All-enabled-disabled} and \textit{statement-coverage} are still on the top position, as they are not sensitive to the number of configuration options.

\subsection{Build-System Information}
\label{sec:bsi}

%LINUX
%Files: 9196.0
%Files with presence condition: 233.0
%Files without presence condition: 8963.0
%Percetual (files with presence condition): 0.02533710308829926
%(1: 301) (2: 3685) (3: 3762) (4: 1102) (5: 183) (6: 53) (7: 28) (8: 12) (9: 27) (10: 2) (11: 1) (15: 24) (17: 1) (16: 12) (20: 1) (25: 2)

%BUSYBOX
%Files: 523.0
%Files with presence condition: 110.0
%Files without presence condition: 413.0
%Percetual (files with presence condition): 0.21032504780114722
%(1: 109) (2: 328) (3: 39) (4: 14) (5: 7) (6: 2) (7: 6) (8: 1) (9: 2) (11: 1) (22: 14)

The build system controls which files are compiled and included.
Files may be included only when specific configuration options are selected or may be compiled
with additional parameters. This is equivalent to wrapping an additional \texttt{\#ifdef} around each source file
or define additional macros in the beginning of a file.
Like ignoring constraints, ignoring build-system information can lead to false positives,
where faults are reported in configurations that are prevented in practice by the build system.

Build systems often have a strong influence on the configurability of a system;
for instance, in the \textit{Linux Kernel}, 97\% of source files are compiled only in
certain configurations, 80\% in \textit{BusyBox}.
Still, extracting configuration knowledge from build systems is very difficult. 
While \textit{Linux} and \textit{Busybox} have been analyzed with specialized parsers
that recognize common patterns~\cite{Berger10,NadiTSE15}, and more modern build systems use a more declarative 
style, which is easier to analyze~\cite{GoogleBazil}, analyzing Make files in general is an open research problem
with only few initial solutions~\cite{TienNguyenICSE12,ZhouReleng14}.

Considering build-system information, the presence conditions of faults become more complex, because we include the condition when the file is compiled: Whereas without build-system information 40\,\% of all faults of our corpus can be found by enabling or disabling a single option, only 17\,\% can be found the same way with build-system information.
By requiring more options to pinpoint faults, incorporating build-system information decreases the efficiency of algorithms.
\textit{Pair-wise}, \textit{three-wise}, \textit{most-enabled-disabled}, and \textit{one-enabled} detected fewer faults than in the first study.
%For instance, \textit{one-enabled} enables only one configuration option at a time, which is not sufficient anymore to detect some faults after adding the file presence conditions to the faults presence conditions.

The sizes of the sample sets are slightly increased in all sampling algorithms (except \textit{most-enabled-disabled}), as we consider additional configuration options used in the build system. Time required to compute sample sets is increased only by a few milliseconds.

%\TODO{summary:} difficult analysis; few more options to consider, but no significant change
\vspace{-0.1cm}
\summary{Including build-system information requires to consider more configuration options in most files, but does not significantly affect any sampling algorithm or their efficiency.}

\section{Threats to Validity}
\label{sec:threats}
Regarding external validity, we studied only systems that implement variability with conditional compilation and cannot generalize to systems that use other mechanisms to implement variability.

%We started our comparative study by making some assumptions (as discussed in Section~\ref{sec:overallsetup}), we excluded faults that span multiple files from our set and considered only faults confirmed by developers.
%This way, we could not study some points about sampling, such as false positives and the capability of the algorithms to detect inter-file faults.
%We and others found many false positives when performing static analysis, because of missing build-system and constraints information~\cite{Tartler11,Medeiros13}.
%In addition, we studied the benefits of global analysis regarding inter-file faults.
%In \textit{Study 2}, we might have faults positives, but we do not have constraint and build-system information to check the warnings reported by \textit{Cppcheck}.

Regarding internal validity, the corpus of faults is critical for our
research strategy.
Creating a representative corpus is difficult, primarily because
we have no means of knowing all faults in the system.
We address this threat with two strategies:
\begin{compactitem}

    \item We avoided biasing our corpus to any specific sampling algorithm. 
    As the corpus has been partially mined from software repositories, 
    it might be biased towards more popular system configurations. 
    Still, our corpus is the most comprehensive corpus of configuration-related faults we are aware of.

    \item We conducted a complimentary experiment using an automated bug-finding technique 
    instead of a corpus of
    known faults. This experiment yielded comparable results
    and complements and confirms the first study on our corpus.
    In a nutshell, we measured with which sampling algorithm 
    the bug-finding technique, static analysis with \textit{Cppcheck}, 
    would expose the most warnings per sampled configuration. 
    The experiment introduces a different threat to internal validity
    in terms of false positives, but triangulating the results across both 
    setups with orthogonal threats increases confidence in our findings.
    We omit details for space reasons and 
    refer the interested reader to Appendix A.
    %\footnote{\url{http://www.dsc.ufcg.edu.br/~spg/appendix.pdf}}
\end{compactitem}

%\vspace{-0.3cm}
\section{Guidance for Practitioners}
\label{sec-guidance}

Our study provides empirical evidence about the efficiency and typical sample sizes for analyzing configurable C code with various sampling algorithms both under ideal and real-world conditions. There is not a single sampling algorithm that is optimal for all systems and in all conditions, but practitioners can use our results to identify plausible candidates for their purposes.

For instance, during initial phases of a project, when developers are changing the source code frequently, they may prefer sampling algorithms with small sample sets to run the analysis fast. At some point, such as before a release, developers might want to use algorithms with larger sample sets, to minimize the number of configuration-related faults.
Under favorable conditions, that is, without considering constraints, global analysis, header files, and build-system information, all algorithms scale in practice; we recommend \textit{t-wise} sampling with a high \textit{t} for rigorous testing and simpler sampling 
algorithms, such as \textit{most-enabled-disabled}, \textit{pair-wise}, \textit{statement-coverage}, and combinations of these, for quicker early runs.
Combining many and expensive sampling strategies does bring only marginal benefits but requires very large sample sets; we do not recommend them.

When considering header files, constraints, and global analysis, we recommend to go for simple algorithms, such as \textit{most-enabled-disabled}, because many other algorithms become intractable in practice, as presented in Table~\ref{fig:summary-dimensions}.

Again, while no algorithm fits all contexts, we hope that our data will help practitioners to identify suitable candidate sampling algorithms for their specific scenario.

\section{Related Work}
\label{s:related}

Several researchers have studied the way developers use the C preprocessor.
They performed empirical studies with open-source systems written in C that are statically configurable with the C preprocessor~\cite{Liebig11,Ernst02,Baxter92}. 
%Hunsen et al.~\cite{Hunsen15} performed a study to understand how the C preprocessor is used in open-source and industrial systems, and
%Passos et al.~\cite{Passos15} studied configuration-related evolution patterns in the \textit{Linux kernel}.
%Schulze et al.~\cite{Schulze13} used a different research method, and performed a controlled experiment to compare disciplined and undisciplined use of preprocessor directives, such as \texttt{\#ifdef} and \texttt{\#endif}.
Liebig et al.~\cite{Liebig11} found that almost 16\% of the preprocessor usage is undisciplined according to an empirical study of 40 C software systems.
In a previous study~\cite{Medeiros15}, we interviewed 40 developers and performed a survey with 202 developers to understand why the C preprocessor is still widely used in practice despite the strong criticism the preprocessor receives in academia.
According to our results, developers check only a few configurations of the source code.
All these studies discussed the C preprocessor and its problems, such as faults, inconsistencies, code quality, and incomplete testing.
These studies motivated us to analyze sampling algorithms to support developers in finding configuration-related faults.

Other studies have analyzed software repositories by considering faults already fixed by developers to understand the characteristics of configuration-related faults~\cite{Medeiros13,Iago14}.
In particular, researchers analyzed configuration-related faults in dynamic configurable systems~\cite{Garvin11-2, Kuhn04}. They concluded that the majority of configuration-related faults involve a few configuration options, a result similar to ours. 
Abal et al.~\cite{Iago14} analyzed the \textit{Linux Kernel} software repository to study con-\\figuration-related faults. 
Tartler et al.~\cite{Tartler14} also performed studies to find configuration-related faults in the \textit{Linux Kernel}.
In our study, we considered some configuration-related faults reported by these previous studies. 
In addition, there are several studies proposing tools to find faults, such as \textit{Undertaker}~\cite{Tartler11-2}, \textit{Tracker}~\cite{Torlak10}, and \textit{Splint}~\cite{Larochelle01}.
%, and \textit{Plug}~\cite{Bond08}.
%However, these tools are not variability-aware and consider one system configuration at a time. 

Researchers have proposed various strategies to deal with configuration-related faults.
They considered combinatorial interaction testing to check different combinations of configuration options and prioritize test cases~\cite{Cohen97theaetg,Kuhn04,Cohen03,Sampath08,Qu08,Yilmaz06}.
Nie et al.~\cite{Nie11} performed a survey with combinatorial testing approaches, but without considering the complexities of C, such as header files and build-system information.
Most studies on sampling make assumptions that might not be realistic in practice, such as ignoring constraints among configuration options.
Including constraints, build-system information, and header files is a non-trivial task.
Several researchers used the \textit{t-wise} sampling algorithm to cover all $t$ configuration option combinations~\cite{Oster10,Perrouin10,Johansen12,Marijan13}, many studies without considering constraints~\cite{Shi05,Kuhn04,Lei08,Nie11}.
%Petke et al.~\cite{Petke15} compared strategies to generate covering arrays for \textit{t-wise} algorithms, such as simulated annealing and greedy algorithms.
Other researchers proposed the \textit{statement-coverage}~\cite{Tartler11-2} sampling algorithm and applied a per-file analysis.
Abal et al.~\cite{Iago14} suggested the \textit{one-disabled} algorithm.
S\'{a}nchez et al.~\cite{AnaB15} studied realistic settings and studied the use of non-functional data for test case prioritization. 
Other researchers applied \textit{t-wise} algorithms with constraints~\cite{Johansen12,Garvin09,Borazjany12}, and Grindal et al.~\cite{Grindal06} studied different constraint handling methods.
However, a comparative study to understand the fault-detection capability and effort (size of sample set) of sampling algorithms, and the influence of limiting assumptions on sampling was not covered in previous studies.

K\"{a}stner et al.~\cite{Kastner11} developed a variability-aware parser that analyzes all possible configurations of a C program simultaneously. It also performs type checking~\cite{Kastner12} and data-flow analysis~\cite{Liebig13}. 
Gazzillo and Grimm~\cite{Gazzillo12} developed a similar parser. 
In our work, we considered faults detected by \textit{TypeChef} and reported in previous studies~\cite{Kastner12-2}. 
Difficulties in setting up these tools and narrow classes of detectable faults limit their applicability and lead to false positives. 
In addition, variability-aware tools work at the preprocessor level, which hinders the reuse of existing bug checkers of traditional C tools, including \textit{Gcc} and \textit{Clang}.

Some studies have compared sample-based and variability-aware strategies. Apel et al.~\cite{Apel13} developed a model checking tool for product lines and used it to compare sample-based and variability-aware strategies with regard to verification performance and the ability to find defects. 
Liebig et al.~\cite{Liebig13} performed studies to detect the strengths and weaknesses of variability-aware and sampling-based analyses. 
They considered two analysis implementations (type checking and liveness analysis) and applied them to a number of subject systems, such as \textit{Busybox} and the \textit{Linux Kernel}.
%Kolesnikov et al.~\cite{Kolesnikov13} compared variability-aware, feature-based, and product-based type checking.
In our study, we performed complimentary analyses regarding sampling algorithms and filled a gap by comparing sampling algorithms considering the influence of assumptions made in previous studies.

\section{Concluding Remarks}
\label{s:conclusion}

We compared 10 sampling algorithms. 
Our study makes a step toward understanding the tradeoffs between effort (i.e., how large are the sample sets) and fault-detection capabilities (i.e., how many faults can be found in the selected configurations).

In a first study, we used a corpus of~\BUGSONE~configuration-related faults from~\PROJECTS~popular C projects reported in previous studies.
We initially ran the comparison accepting some assumptions and we ignored configuration constraints, header files and build-system information, and we applied a per-file analysis.
The results reveal that all sampling algorithms selected configurations that include at least 66\% of the~\BUGSONE~faults reported in previous work. As expected, the algorithms with the largest sample sizes detected the most faults. More interestingly, we identified several combinations of algorithms that provide a useful balance between sample size and fault-detection capabilities.

Subsequently, we performed a complementary study to measure the influence of considering constraints, global analysis, header files, and build-system information on sampling.
We found that, when considering constraints, we can reduce false positives, but it increases the costs for generating sample sets, which are often not optimal.
Using a global analysis, we can potentially detect non-modular faults that span multiple files, but it causes an explosion in the number of considered configuration options that leads to large sample sets.
When incorporating header files, there is a potential to detect additional faults, but the setup is difficult and the algorithms produce much larger sample sets.
When including build-system information, we face a difficult analysis, a few more configuration options to consider, but no significant changes.
Overall, we found that global analysis and analyses that include configuration options from header files turn the analysis to be practically infeasible for most algorithms.

Our study fills a gap, as a comparison of sampling algorithms for finding configuration-related faults was not available.
Our findings are meant to support developers in understanding the tradeoffs regarding effort and fault-detection capabilities of sampling algorithms, aiding developers in selecting an algorithm and deciding what kind of information they include in the analysis.
A lack of understanding the tradeoffs and assumptions of sampling algorithms can lead to both, undetected faults, which decreases software quality, and time-consuming code analysis, which increases costs.

\section*{Acknowledgments}

This work has been supported by CNPq 460883/2014-3, 573964/2008-4 (INES), 477943/2013-6, and 306610/2013-2, CAPES 175956, project DEVASSES (European Union Seventh Framework Programme, agreement no PIRSES-GA-2013-612569), NSF award 1318808, the U.S. Department of Defense through the Systems Engineering Research Center (SERC, contract H98230-08-D-0171), the Science of Security Initiative, and the German Research Foundation (AP 206/4 and AP 206/6).

\newpage
\bibliographystyle{abbrv}
\bibliography{references}  % sigproc.bib is the name of the Bibliography in this case
% ACM needs 'a single self-contained file'!
%\balancecolumns
\newpage
\appendix
%Appendix A
\section{Cppcheck Warnings}
\label{s:evaluation}

The goal of this experiment is to compare the sampling algorithms (\textbf{RQ1--3}) from a different perspective. Instead of measuring fault-detection capabilities in terms of a corpus of known configuration-related faults, we use a static-analysis tool as our automated fault-detection mechanism.

They key difference to \textit{Study 1} is how we operationalize the dependent variable regarding fault-detection capabilities.
Unfortunately, there is no tool that would produce a reliable ground truth.\footnote{\scriptsize 
	Variability-aware analysis tools, such as \textit{TypeChef}~\cite{Kastner11,Kastner12-2} and \textit{SuperC}~\cite{Gazzillo12},
	could soundly cover all configurations regarding syntax or type errors, but would require a time-consuming initial setup that would make our study infeasible.
}
We run the static analysis tool (\textit{Cppcheck}) on each sampled configuration of each file and count all reported warnings.
We discard warnings that occur in all configurations, because they are not configuration related.
Although a warning does not necessarily correspond to a fault, it provides a rough estimate of the number of issues a
developer would have to investigate (also \textit{Cppcheck} claims to minimize false positives). 
We assume that the distribution of warnings throughout the code is roughly similar to the distribution
of real faults in C code and can hence serve as a proxy to measure how configuration-related faults are distributed
over the configuration space.

We performed the study on a fresh set of subject systems, that does not overlap with the corpus of \textit{Study 1}:
\textit{expat},
\textit{flex},
\textit{gimp},
\textit{gnumeric},
\textit{gzip},
\textit{kindb},
\textit{mplayer},
\textit{mpsolve},
\textit{mptris},
\textit{openldap},
\textit{parrot},
\textit{prc-tools},
\textit{privoxy},
\textit{sylpheed},
\textit{tk},
\textit{xine-lib}.
We selected these systems guided by previous work~\cite{Liebig11,Liebig10}, which identified projects statically configurable with the C preprocessor.

\subsection{Results and Discussion}
\label{operation}
\begin{figure}[ht]
\centering
\vspace{-0.2cm}
\includegraphics[width=76mm]{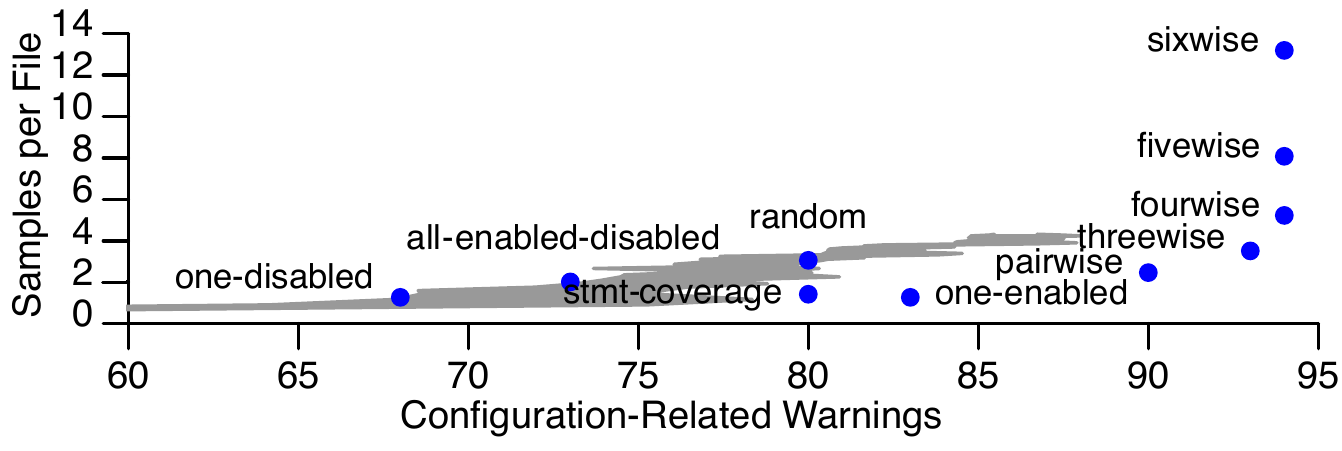}
\footnotesize{
\begin{tabular}{p{4.5cm} c c }
  \\
   \textbf{Sampling Algorithm} & \textbf{Faults} & \textbf{Samples}\\
    \hline
  One-disabled & 68 & 1.3 \\
  Most-enabled-disabled & 73 & 2.0 \\
  Statement-coverage & 80 & 1.4 \\
  Random & 80 & 3.1 \\
  One-enabled & 83 & 1.3 \\
  Pair-wise & 90 & 2.5 \\
  Three-wise & 93 & 3.5 \\
  Four-wise & 94 & 5.2 \\
  Five-wise & 94 & 8.1 \\
  Six-wise & 94 & 13.2 \\
  \hline
\end{tabular}
}
\caption{Number of warnings reported and samples per file for each sampling algorithm considered in \textit{Study 2}.}
\vspace{-0.2cm}
\label{fig:sampling-analysis-real}
\end{figure}

\begin{figure}[ht]
\centering
\vspace{-0.2cm}
\includegraphics[width=76mm]{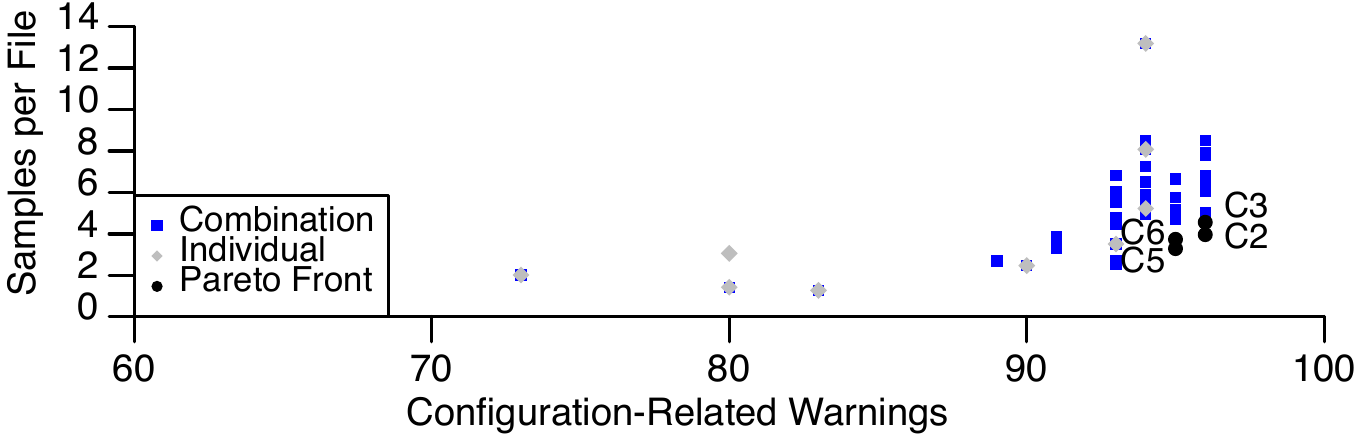}
\footnotesize{
\begin{tabular}{p{8cm}}
   \\
   \textbf{Sampling Algorithm}\\
    \hline
    C2 One-enabled, one-disabled and statement-coverage\\
    C3 One-enabled, one-disabled and most-enabled-disabled\\
    C5 One-enabled, and most-enabled-disabled \\
    C6 Pair-wise and one-enabled\\
  \hline
\end{tabular}
\begin{tabular}{p{0.4cm} c c p{0.4cm} c c}
   \\
   \textbf{} & \textbf{Faults} & \textbf{Samples} & \textbf{} & \textbf{Faults} & \textbf{Samples}\\
    \hline
    C2  & 96 & 4.0 & C5 & 95 & 3.3 \\
    C3  & 96 & 4.6 & C6 & 95 & 3.7\\
  \hline
\end{tabular}
}
\caption{Number of faults and samples per file for the combinations of sampling algorithms on the \textit{Pareto Front}.}
\vspace{-0.2cm}
\label{fig:sampling-cppcheck-combinations}
\end{figure}

Overall, \textit{Cppcheck} reported 96 warnings that appear only in specific configurations of the code over 77 distinct files.
All 10 sampling algorithms reported more than 70\% of the 96 configuration-related warnings, and no sampling algorithm reported all 96 warnings.
We summarize the results of this experiment in Figure~\ref{fig:sampling-analysis-real}.
Again, \textit{five-wise} and \textit{six-wise} reported the highest number of warnings.
\textit{One-disabled} and \textit{statement-coverage} reported the lowest number of warnings.
There is a warning for \textit{Xine-lib}, where developers need to disable eight distinct configuration options to make \textit{Cppcheck} report it. 
\textit{Six-wise} misses this specific warning.
However, other sampling algorithms, such as \textit{most-enabled-disabled} and \textit{one-enabled}, reported the 
warning for \textit{Xine-lib}.
Furthermore, we computed the number of warnings reported for the combinations of sampling algorithms and found combinations that
reported all 96 warnings (e.g., \textit{C2} and \textit{C3}), as depicted in Figure~\ref{fig:sampling-cppcheck-combinations}.

The sizes of sample sets range from 1.3 to 13.2 configurations per file.
Again, \textit{six-wise} selected the highest number of configurations (more than 100K across all projects), while \textit{one-enabled} and \textit{one-disabled} selected the lowest number of configurations.
The majority of the combinations of algorithms created a very large sample set.
Figure~\ref{fig:sampling-cppcheck-combinations} presents four combinations of sampling algorithms on the \textit{Pareto Front}: \textit{C2}, \textit{C3}, \textit{C5}, and \textit{C6}.

We computed the ranking of algorithms considering the efficiency function of Section~\ref{sec-study1-results}. 
The algorithms, starting from the most efficient, are: \textit{one-enabled},
\textit{stmt-coverage},
\textit{one-disabled},
\textit{pair-wise},
\textit{most-enabled-disabled},
\textit{three-wise},
\textit{random},
\textit{four-wise},
\textit{five-wise}, and 
\textit{six-wise}. 
Overall, the ranking is stable when compared to \textit{Study 1} and there were only minor changes: \textit{most-enabled-disabled} and \textit{pair-wise} are less efficient here, while
\textit{one-enabled}, \textit{one-disabled}, and \textit{statement-coverage} are more efficient.
These changes can be explained by analyzing the number of files with only one configuration option, which is higher in our experiment than in \textit{Study 1}. 
\textit{Most-enabled-disabled} requires two configurations for each file with one configuration option; \textit{one-enabled} and \textit{one-disabled} require only one configuration per file. 
It makes \textit{one-enabled} and \textit{one-disabled} more efficient and impacts the ranking.
Regarding the five least efficient algorithms, the ranking is exactly the same as in \textit{Study 1}.

\textit{Study 1} and this experiment complement and confirm each other, as we obtain essentially the same results 
regarding the fault-detection capabilities of the sampling algorithms by using different perspectives: known faults reported in previous studies (\textit{Study 1}) and \textit{Cppcheck} as our fault-detected mechanism.
We found two combinations of sampling algorithms (\textit{C2}, and \textit{C3}) that are on the \textit{Pareto Front} of \textit{Study 1} as well, which support them as efficient combinations.
By triangulating the results, we gain confidence in the findings.

\end{document}